\documentclass[twocolumn,twocolappendix]{aastex631}
\usepackage{amsmath}
\usepackage{amssymb}
\usepackage{multirow}
\usepackage{tablefootnote}
\usepackage[T5]{fontenc}

\newcommand{\fesc}{\ifmmode{f_{\rm esc}}\else{$f_{\rm esc}$}\fi}
\newcommand{\fescs}{\ifmmode{f_{\rm esc}^\star}\else{$f_{\rm esc}^\star$}\fi}
\newcommand{\kms}{\ifmmode{{\;\rm km~s^{-1}}}\else{km~s$^{-1}$}\fi}
\newcommand{\fgas}{\ifmmode{{f_{\rm gas}}}\else{$f_{\rm gas}$}\fi}
\newcommand{\cubecm}{\ifmmode{{\rm cm^{-3}}}\else{cm$^{-3}$}\fi}
\newcommand{\ztwo}{\ifmmode{{\rm [Z_2/H]}}\else{[Z$_2$/H]}\fi}
\newcommand{\zthree}{\ifmmode{{\rm [Z_3/H]}}\else{[Z$_3$/H]}\fi}
\newcommand{\lsim}{\lower0.3em\hbox{$\,\buildrel <\over\sim\,$}}
\newcommand{\gsim}{\lower0.3em\hbox{$\,\buildrel >\over\sim\,$}}

\newcommand{\sfr}{\ifmmode{\textrm{M}_\odot \,\textrm{yr}^{-1} \,\textrm{Mpc}^{-3}}\else{M$_\odot$ yr$^{-1}$ Mpc$^{-3}$}\fi}
\newcommand{\hsfr}{\ifmmode{\textrm{M}_\odot\, \textrm{yr}^{-1}}\else{M$_\odot$ yr$^{-1}$}\fi}

\newcommand{\eavg}{\ifmmode{\langle E_\gamma \rangle}\else{$\langle E_\gamma \rangle$}\fi}

\newcommand{\Ms}{\ifmmode{M_\odot}\else{$M_\odot$}\fi}
\newcommand{\vrms}{\ifmmode{v_{\rm rms}}\else{$v_{\rm rms}$}\fi}

\newcommand{\tvir}{\ifmmode{T_{\rm{vir}}}\else{$T_{\rm{vir}}$}\fi}
\newcommand{\mvir}{\ifmmode{M_{\rm{vir}}}\else{$M_{\rm{vir}}$}\fi}
\newcommand{\rvir}{\ifmmode{r_{\rm{vir}}}\else{$r_{\rm{vir}}$}\fi}

\newcommand{\jj}{\ifmmode{J_{21}}\else{$J_{21}$}\fi}
\newcommand{\flw}{\ifmmode{F_{LW}}\else{$F_{LW}$}\fi}
\newcommand{\kph}{\ifmmode{k_{\rm ph}}\else{$k_{\rm ph}$}\fi}

\newcommand{\zsun}{\ifmmode{\rm\,Z_\odot}\else{$\rm\,Z_\odot$}\fi}

\newcommand{\nhi}{\ifmmode{N_{\rm HI}}\else{$N_{\rm HI}$}\fi}

\begin{document}

\title{The AGORA High-resolution Galaxy Simulations Comparison Project. XI: Solving the Non-Spherical Morphology and Evolution of Dark Matter Halos with {\sc Haskap Pie}}

\author[0000-0002-8638-1697]{Kirk S.~S.~Barrow}
\affiliation{Department of Astronomy, University of Illinois Urbana-Champaign,
1002 W Green St, Urbana, IL 61801, USA; \rm{\href{mailto:kbarrow@illinois.edu}{kbarrow@illinois.edu}}}

\author[0009-0002-2290-8039]{Thịnh Hữu Nguyễn}
\affiliation{Department of Astronomy, University of Illinois Urbana-Champaign,
1002 W Green St, Urbana, IL 61801, USA}

\author[0000-0002-6299-152X]{Santi Roca-F\`{a}brega}
\altaffiliation{Code leaders}
\affil{Lund Observatory, Division of Astrophysics, Department of Physics, Lund University, SE-221 00 Lund, Sweden}
\affil{Departamento de F\'{i}sica de la Tierra y Astrof\'{i}sica, Facultad de Ciencias F\'{i}sicas, Plaza Ciencias, 1, 28040 Madrid, Spain}

\author[0000-0003-4464-1160]{Ji-hoon Kim}
\altaffiliation{Code leaders}
\affiliation{Seoul National University Astronomy Research Center, Seoul 08826, Korea}
\affiliation{Center for Theoretical Physics, Department of Physics and Astronomy, Seoul National University, Seoul 08826, Korea}
\affiliation{Institute for Data Innovation in Science, Seoul National University, Seoul 08826, Republic of Korea}

\author[0009-0009-7224-4462]{Varun Satish}
\affiliation{Department of Astronomy, University of Illinois Urbana-Champaign,
1002 W Green St, Urbana, IL 61801, USA}

\author[0000-0001-7457-8487]{Kentaro Nagamine}
\altaffiliation{Code leaders}
\affiliation{Theoretical Astrophysics, Department of Earth and Space Science, Graduate School of Science, Osaka University, Toyonaka, Osaka, 560-0043, Japan}
\affiliation{Theoretical Joint Research, Forefront Research Center, Graduate School of Science, The University of Osaka, 1-1 Machikaneyama, Toyonaka, Osaka 560-0043, Japan}
\affiliation{Kavli IPMU (WPI), UTIAS, The University of Tokyo, Kashiwa, Chiba 277-8583, Japan}
\affiliation{Department of Physics \& Astronomy, University of Nevada Las Vegas, Las Vegas, NV 89154, USA}
\affiliation{Nevada Center for Astrophysics, University of Nevada, Las Vegas, 4505 S. Maryland Pkwy, Las Vegas, NV 89154-4002, USA}

\author[0009-0005-5218-2461]{Saulius Matusaitis}
\affiliation{Department of Astronomy, University of Illinois Urbana-Champaign,
1002 W Green St, Urbana, IL 61801, USA}

\author[0009-0000-2762-7987]{Eduárd Illes}
\affiliation{Kapteyn Astronomical Institute, University of Groningen, Landleven 12, NL-9747 AD Groningen, the Netherlands}
\affiliation{Tartu Observatory, University of Tartu, Observatooriumi 1, 61602 Tõravere, Estonia}

\author[0000-0002-9158-195X]{Ram\'{o}n Rodr\'{i}guez-Cardoso}
\affil{Departamento de Física de la Tierra y Astrofísica, Fac. de C.C. Físicas, Universidad Complutense de Madrid, E-28040 Madrid, Spain}
\affil{GMV, Space and Avionics Equipment, Isaac Newton, 11 Tres Cantos, E-28760 Madrid, Spain}
\affil{Instituto de Física de Partículas y del Cosmos, IPARCOS, Fac. C.C. Físicas, Universidad Complutense de Madrid, E-28040 Madrid, Spain}

\author[0000-0002-9144-1383]{Minyong Jung}
\affiliation{Center for Theoretical Physics, Department of Physics and Astronomy, Seoul National University, Seoul 08826, Korea}

\author[0000-0002-7820-2281]{Hyeonyong Kim}
\altaffiliation{Code leaders}
\affiliation{Center for Theoretical Physics, Department of Physics and Astronomy, Seoul National University, Seoul 08826, Korea}

\author{Anna Genina}
\altaffiliation{Code leaders}
\affil{Institute for Astronomy, University of Edinburgh, Royal Observatory, Blackford Hill, Edinburgh EH9 3HJ, UK}

\author[0009-0002-1398-6537]{Pablo Granizo}
\affiliation{Theoretical Astrophysics, Department of Earth and Space Science, Graduate School of Science, Osaka University, Toyonaka, Osaka, 560-0043, Japan}
\affil{Universidad Aut\'{o}noma de Madrid, Ciudad Universitaria de Cantoblanco, E-28049 Madrid, Spain}

\author[0000-0001-6106-7821]{Alessandro Lupi}
\altaffiliation{Code leaders}
\affil{Como Lake Center for Astrophysics, DiSAT, Universit\`a degli Studi dell'Insubria, via Valleggio 11, I-22100 Como, Italy}
\affil{INFN, Sezione di Milano-Bicocca, Piazza della Scienza 3, I-20126 Milano, Italy}
\affil{INAF - Osservatorio di Astrofisica e Scienza dello Spazio di Bologna, Via Gobetti 93/3, I-40129 Bologna}

\author[0000-0002-3764-2395]{Johnny W. Powell}
\altaffiliation{Code leaders}
\affil{Department of Physics, Reed College, Portland, OR 97202, USA}

\author{H\'{e}ctor Vel\'{a}zquez}
\altaffiliation{Code leaders}
\affil{Instituto de Astronom\'{i}a, Universidad Nacional Aut\'{o}noma de M\'{e}xico, A.P. 70-264, 04510, Mexico, D.F., Mexico}

\author[0000-0002-5969-1251]{Tom Abel}
\affil{Kavli Institute for Particle Astrophysics and Cosmology, Stanford University, Stanford, CA 94305, USA}
\affil{Department of Physics, Stanford University, Stanford, CA 94305, USA}
\affil{SLAC National Accelerator Laboratory, Menlo Park, CA 94025, USA}

\author[0000-0002-4287-1088]{Oscar Agertz}
\affil{Lund Observatory, Division of Astrophysics, Department of Physics, Lund University, SE-221 00 Lund, Sweden}

\author[0000-0001-8531-9536]{Renyue Cen}
\affil{Center for Cosmology and Computational Astrophysics, Institute for Advanced Study in Physics, Zhejiang University, Hangzhou 310027, People's Republic of China}
\affil{Institute of Astronomy, School of Physics, Zhejiang University, Hangzhou 310027, People's Republic of China}

\author[0000-0002-8680-248X]{Daniel Ceverino}
\affil{Universidad Aut\'{o}noma de Madrid, Ciudad Universitaria de Cantoblanco, E-28049 Madrid, Spain}
\affil{CIAFF, Facultad de Ciencias, Universidad Aut\'{o}noma de Madrid, E-28049 Madrid, Spain}

\author[0000-0003-4597-6739]{Boon Kiat Oh}
\affiliation{Department of Physics, University of Connecticut, U-3046, Storrs, CT 06269, USA}
\affiliation{School of Physics, Korea Institute for Advanced Study, 85 Hoegiro, Dongdaemun-gu, Seoul 02455, Republic of Korea}

\author[0000-0002-5712-6865]{Yuri Oku}
\affil{Center for Cosmology and Computational Astrophysics, Institute for Advanced Study in Physics, Zhejiang University, Hangzhou 310027, People's Republic of China}

\author[0000-0001-5091-5098]{Joel R. Primack}
\altaffiliation{In memoriam}
\affil{Department of Physics, University of California at Santa Cruz, Santa Cruz, CA 95064, USA}

\author[0000-0001-5510-2803]{Thomas R. Quinn}
\affil{Department of Astronomy, University of Washington, Seattle, WA 98195, USA}

\author{Yves Revaz}
\altaffiliation{Code leaders}
\affil{Institute of Physics, Laboratoire d'Astrophysique, \'{E}cole Polytechnique F\'{e}d\'{e}rale de Lausanne (EPFL), CH-1015 Lausanne, Switzerland}

\author[0000-0002-0415-3077]{Alvaro Segovia-Otero}
\affil{Lund Observatory, Division of Astrophysics, Department of Physics, Lund University, SE-221 00 Lund, Sweden}

\author{Ikkoh Shimizu}
\altaffiliation{Code leaders}
\affil{Shikoku Gakuin University, 3-2-1 Bunkyocho, Zentsuji, Kagawa, 765-8505, Japan}

\author[0009-0005-5945-209X]{Edward Skrabacz}
\affiliation{Department of Astronomy, University of Illinois Urbana-Champaign,
1002 W Green St, Urbana, IL 61801, USA}

\author{Romain Teyssier}
\affil{Department of Astrophysical Sciences, Princeton University, Princeton, NJ 08544, USA}

\author{the {\it AGORA} Collaboration}
\footnote{The list of authors is provisional and is just in alphabetical order. Before the beta release of this paper, the order will be discussed with all the coauthors and a final list will be included.}


\begin{abstract}

We introduce a halo solving and tracking procedure that intrinsically treats dark matter halos as non-spherical objects by leveraging the bound particle searching techniques used in {\sc Haskap Pie} \citep{2026ApJ...999...72B}. The \textit{AGORA} Collaboration's hydrodynamic simulation \texttt{CosmoRun} project provides a useful laboratory to explore trends in dark matter halo morphology that are revealed by our new procedure in the context of any dispersions or similarities between the codes. We find that several morphological and shape measures were very responsive to high mass ratio mergers. The greatest difference in these measures between the simulation codes were related to timing discrepancies and the dynamical state of the halos prior to the mergers. Most other quantities were similar across codes, including several secular and redshift-dependent trends in various dynamical quantities that showed a departure from Virial Theorem (e.g., overdensity and halo mass). We find that halo spin and the ratio between the semi-major and the semi-minor axis peaked at $4>z>2$ before declining at low redshift. Also, halo overdensity is both mass-dependent and redshift-dependent, diverging for low mass halos at low redshift. Our method contributes a new perspective on these trends that have not been fully replicated in other works due to our emphasis on fundamentally non-spherical halos and measures of morphology that correspondingly do not assume spherical symmetry.

\end{abstract} 

\section{Introduction}

For some time, the study of dark matter halos has focused on a model of spherical collapse. Therein, the Universe begins as a dense, expanding collection of small perturbations in matter density. Expansion is rapid enough that matter cannot immediately collapse under gravity. However, after a period of growth of scale, the local gravitational potential of local density peaks leads to a bias in the density field of matter. If this density reaches $\sim$1.69 times the critical density of the Universe \citep{1972ApJ...176....1G}, self-attraction eventually overcomes cosmological expansion, and the matter begins to contract under gravity. Eventually, this leads to runaway growth of density, and matter enters gravitational free-fall, collecting into halos.

If one assumes that the matter falling into a halo is spherically symmetric and has uniform density, this process can be modeled with parametric equations for gravitational acceleration \citep{1965ApJ...142.1431L}. Typically, one applies a virialization argument, assuming that the potential energy and total energy of the particles are connected by a scalar factor, $n$, that depends on the density profile of the halo \citep{1991MNRAS.251..128L}. From here, the final radius can be determined based on the cosmological time of the start of collapse and the time of virialization. In the simplest case, the virialization time is assumed to be twice as long as the time of the start of collapse due to the cyclical nature of the solution. In this formalism, if $n=1$, the resulting density of the halo is $\Delta_c=\rho/\bar{rho}= 18\pi^2$ times the critical density of the Universe, $\bar{\rho} = \rho_c = \frac{3H^2}{8\pi G}$ \citep{1980lssu.book.....P}. In several works, an approximated value of $\Delta_c=200$ is used.

This initial matter-only derivation was expanded by \citet{1998ApJ...495...80B} to include expressions for dark energy to form the widely, but not universally, accepted definition of a virial overdensity given by

\begin{equation}
\Delta_c = 18\pi^2+82x-39x^2.
\label{eq: BN}
\end{equation}

Here, $x$ is $\Omega_m (1+z)^3/E^2(z)-1$, where $\Omega_m$ and $E(z)$, where $E(z) = H(z)/H_o$ take their fiducial cosmological definitions. These coefficients were discovered by fitting parameters to analytical solutions, which were based on a reformulated spherical collapse expression that included both matter and dark energy: $\Delta_c = \Omega(a_c)(a_c/r_c)^3$ \citep{1996MNRAS.282..263E}. Here, $a_c$ is the scale factor at collapse and $r_c$ is the radius of the uniform density volume at collapse. Thus, in its development, Eq. \ref{eq: BN}  never breaks from the presumption that halos are spheres or that the medium that forms a halo is homogeneous and spherically symmetric. However, even early analysis of the shape of halos in cosmological simulations showed that they could diverge significantly from a sphere, especially during mergers \citep[e.g.][]{1987ApJ...319..575B,1988ApJ...327..507F,1994MNRAS.271..676L}.

Studies on the non-spherical shapes of halos have heavily focused on the establishment of ellipsoidal semi-major, semi-minor, and intermediate axes and studying the distribution and evolution of ratios of these axes \citep[e.g.][]{2002sgdh.conf..109B,2004IAUS..220..421S,2005ApJ...627..647B,2006MNRAS.367.1781A,2015MNRAS.449.3171B,2016MNRAS.458.4477T,2017MNRAS.467.3226V}. In each case they have found that a significant fraction of halos departs from a spherical shape. Recent observational studies of the Milky Way have also lent evidence to non-spherical halos \citep[e.g.][]{2025A&A...703A..43Z,2022AJ....164..249H,2026MNRAS.tmp..229D}.

When examining dark matter halos in simulations, the presumption of spherical symmetry can also propagate through to the results through the use of a halo-finder. In establishing the halo population and characterizing its physical dimensions, most morphology studies of cosmological simulations have relied on either spherical overdensity (SO) \citep{1994MNRAS.271..676L,1998ApJ...498..137E,2009ApJS..182..608K,2010A&A...519A..94P,2022MNRAS.509..501H,2022A&A...664A..42V} or friends-of-friends(FoF)/phase-space halo solvers \citep[e.g.][]{1985ApJ...292..371D,2001MNRAS.328..726S,2009MNRAS.399..497D,2010ApJS..191...43S,2006ApJ...649....1D,2009MNRAS.396.1329M,2019PASA...36...21E,Behroozi_2013}. SO solvers typically solve a density field to find a halo center and decide dark matter particle membership based on spherical assumptions for overdensity and the arrangement of the density field. FoF/phase-space solvers allow for linking particles without necessarily assuming spherical profiles, but do presume the existence of a halo radius and may unbind particles in that radius after establishing a spherical halo. In both cases, there can be less tolerance for highly non-spherical halos or halos that do not have well-defined virial radii. Therefore, the preponderance of morphological studies has relied on the presumption of spherical halos either in the construction of their measures or in the pipeline used to identify and characterize the halos to some degree.

In addition to studies of ellipsoidal axes, several studies have explored density profiles of halos to explore their internal morphology and confirm the scaling used to derive their virial overdensity. Based on the premise of a spherically symmetric halo, \citet{1996ApJ...462..563N} generated a two-parameter density profile model (NFW) by fitting results from N-body simulations. This model or the earlier \citet{1965TrAlm...5...87E} model have been extensively employed to construct models of dark matter halos in various applications with great success with over 9,000 citations combined. Measures like halo concentration, which come directly from the parameters of the NFW profile and derived measures like \citet{2001ApJ...555..240B} spin (over 1,100 citations), have been extensively modeled, studied, and explored. These profiles and any derived quantities assume that halos have a single density peak in addition to the assumptions of spherical symmetry. This treatment has been broadly successful in predicting and describing the attributes of halos and structure in a wide variety of applications. However, at least in the case of major mergers, halos do not have a spherical shape or a single density peak, which implies that the more active a halo merger tree, the less these assumptions might apply.

This tension has established a need to reexamine the treatment of merging halos and subhalos in halo-finders to produce dedicated halo tracking algorithms that extend their results to better track infalling and subhalos through the dynamical disruptions that warp their characteristics \citep[e.g.][]{2024ApJ...970..178M,2024MNRAS.533.3811D,2025A&A...698A.303R,2025arXiv250310766K}. {\sc Haskap Pie} (Halo finding Algorithm with efficient Sampling, K-means clustering, tree-Assembly, Particle tracking, Python modules, Inter-code applicability, and Energy solving) was similarly developed to study merging and infalling halos, but was built from the ground up to incorporate techniques that do not rely on the establishment of a density peak, a linking length, or a spherical overdensity. Instead, {\sc Haskap Pie} finds halos based on searching the volume self-bound collections of particles, which leads to more dynamically consistent halos that can be tracked in deep and complex potential wells \citep{2026ApJ...999...72B}.

In this work, we deploy two tools to reexamine the spherical halo paradigm to extend studies of halo morphology into potentially new regimes. First, we take advantage of the open-ended halo-solving strategy in {\sc Haskap Pie} to eliminate the assumption of spherical symmetry in the characterization of halos, which was first explored in was first examined in \citet{2026MNRAS.545f2045N}. The primary goal of this study is to present this change and explore of how shape variables can be constructed consistently from non-spherical halos. 

The second tool we leverage is the nine hydrodynamical simulation code-spanning \textit{AGORA} \texttt{CosmoRun} comparison project \citep{2021ApJ...917...64R,2024ApJ...968..125R}, which we used to test our procedure and study halo morphology. \texttt{CosmoRun} is powerful for this study because of the careful way the prescriptions were well converged across simulations codes so well that differences in the results can be directly tied to a few differences in feedback, for example. In short order, the \texttt{CosmoRun} dataset has been the subject of several extensive investigations into various regimes and effects including the properties and populations of satellite galaxies \citep{2024ApJ...964..123J}, the relationship between feedback and the chemistry of the circumgalactic medium \citep{2024ApJ...962...29S}, the dynamical and tidal stripping of satellites \citep{2025A&A...698A.303R}, and the morphology of the stellar and gas disk \citep{2025ApJ...994..245J}. Since baryon effects are well-studied and increasingly constrained for \texttt{CosmoRun}, we can study dark matter halos with a good context for why any differences between simulations may occur.

The structure of this paper is as follows: We begin with a discussion of the modifications to {\sc Haskap Pie} followed by a description of our shape parameters and how they trend for a single snapshot in Sec. \ref{sec: methods}. The primary results (Sec. \ref{sec: results}) are split into two sections, where we first explore the evolution of a single Milky Way-sized main halo across simulations in Sec. \ref{sec: mainhalo} and then the evolution of the whole sample in Sec. \ref{sec: fullsample}. In Sec. \ref{sec: mass-fixed}, we narrow down our sample to a single mass range to explore and explain new trends that will need further context beyond the \textit{AGORA} simulations. Finally, we discuss and summarize our findings in Sec. \ref{sec: summary}.

\section{Methods}

\label{sec: methods}

\subsection{Simulations}

The \textit{AGORA} \texttt{CosmoRun} sample now includes nine codes that span different hydrodynamic solving schemes including adaptive mesh refinement (AMR; ART-I \citep{Kravtsov+1997}, ENZO\citep{Bryan2014}, and RAMSES \citep{Teyssier+2002}), smoothed particle hydrodynamics (SPH; CHANGA \citep{Jetley+2008a, Jetley+2010, Menon+2015}, GADGET-3 as well as GADGET-4 \citep{Springel+2005,Springel+2021} and GEAR \citep{Revaz+2012}), and hybrid methods (AREPO \citep{Springel+2010, Weinberger+2020} and GIZMO \citep{Hopkins+2015}) run with common initial conditions and prescriptions. Each simulation follows a zoom-in region centered on a $10^{12}$ M$_\odot$ halo at $z=0$ with a dark matter mass resolution of $\sim2.8 \times 10^5$ M$_\odot$, as described in \citet{2021ApJ...917...64R,2024ApJ...968..125R}. The public release of the data is described in \citet{2024arXiv240800432R}. CHANGA and GEAR had a large number of timesteps available for analysis, so we took every second and third timestep for those simulations, respectively, to better homogenize the data sampling frequency. For AREPO, we use thermal feedback (AREPO-T). For CHANGA, we use thermal and kinetic feedback rather than superbubbles.

In addition, Fig. \ref{fig: comp1} and Fig. \ref{fig: hull1} were made with an ENZO high-redshift radiative-hydrodynamic zoom-in simulation described in \citet{2024ApJ...969..144S} and \citet{2026ApJ...999...72B} with an effective dark matter mass resolution of $2.81 \times 10^4$ M$_\odot$ centered around a $\sim 1.3 \times 10^9$ M$_\odot$ halo at $z=7.5$. This simulation was used to quickly iterate on changes to the halo-finding code.

\subsubsection{Refined Region Contamination}

Since all our simulations are zoom-in regions that focus on the Lagrangian region of a single halo, it is important that all the halos in our analysis are largely unaffected by contamination from larger particle masses. In \citealt{2026ApJ...999...72B}(Section 2.1.1), we described our refined region solving algorithm, which is distinct from the restrictions built into the simulations. For each timestep of each simulation, we define a bounding box that only includes the most refined particles and second-most refined particles (eight times the mass of the most refined particles), and we excluded all halos outside that region. Overall, less than 5\% of halos in this study volume had any second-most refined particles.

\subsection{Non-Spherical Halo-Solving}

\label{sec: non-sphere}

In prior work \citep{2026ApJ...999...72B}, we used spherical overdensity and gravitational boundedness as a proxy to determine which dark matter particles were part of halos. This method carried the underlying assumption that halos are roughly spherical, which did not leverage a long heritage of studies showing that halos are better described triaxially or through other non-spherical descriptions. In this work we attempt to define non-spherical halos in a way that allows us to study their intrinsic morphology. Our method differs in that we are able to fully describe the potential well of our halos through our particle sampling technique, which allows us to directly determine the shape of the halo during its identification in the simulation. The details of our particle sampling technique are extensively described in the antecedent paper, but in summary, particles are grouped into HealPix-based\citep{2005ApJ...622..759G} annular sections in such a way that the radial distribution of particle number density is roughly flat rather than cuspy. Then the mass of particles in each annular sector is correspondingly upscaled so that the mass distribution is preserved. To find which of these particles are bound to a halo, we segregate particles based on their position, kinetic, and potential energies with respect to the center of a search region using k-means clustering and use the clusters. We then re-solve the energies within each cluster twice, using the bound particles from the previous iteration to define the centers used for potential and kinetic energy calculations.

\begin{figure*}
\begin{center}
\includegraphics[width=\linewidth]{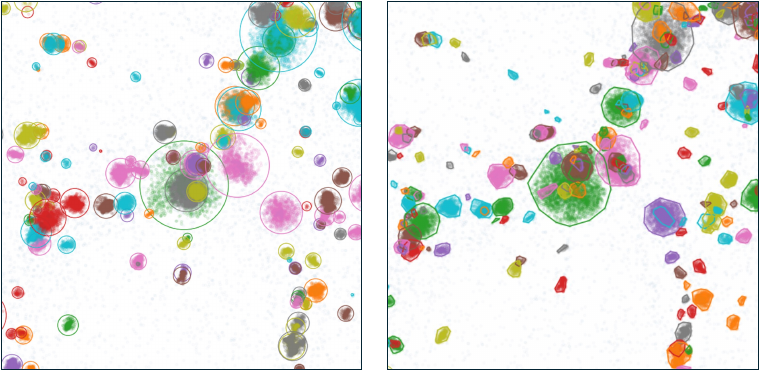}
\caption{Comparison of corresponding results on our test simulation using spherical (left) and non-spherical (right) halo-finding techniques. Particles that are bound to the halo are plotted with the same color as the halo radius. Colors are consistent by halo but rotate through the pallette between halos. }
\label{fig: comp1}
\end{center}
\end{figure*}

\begin{figure}
\begin{center}
\includegraphics[width=\linewidth]{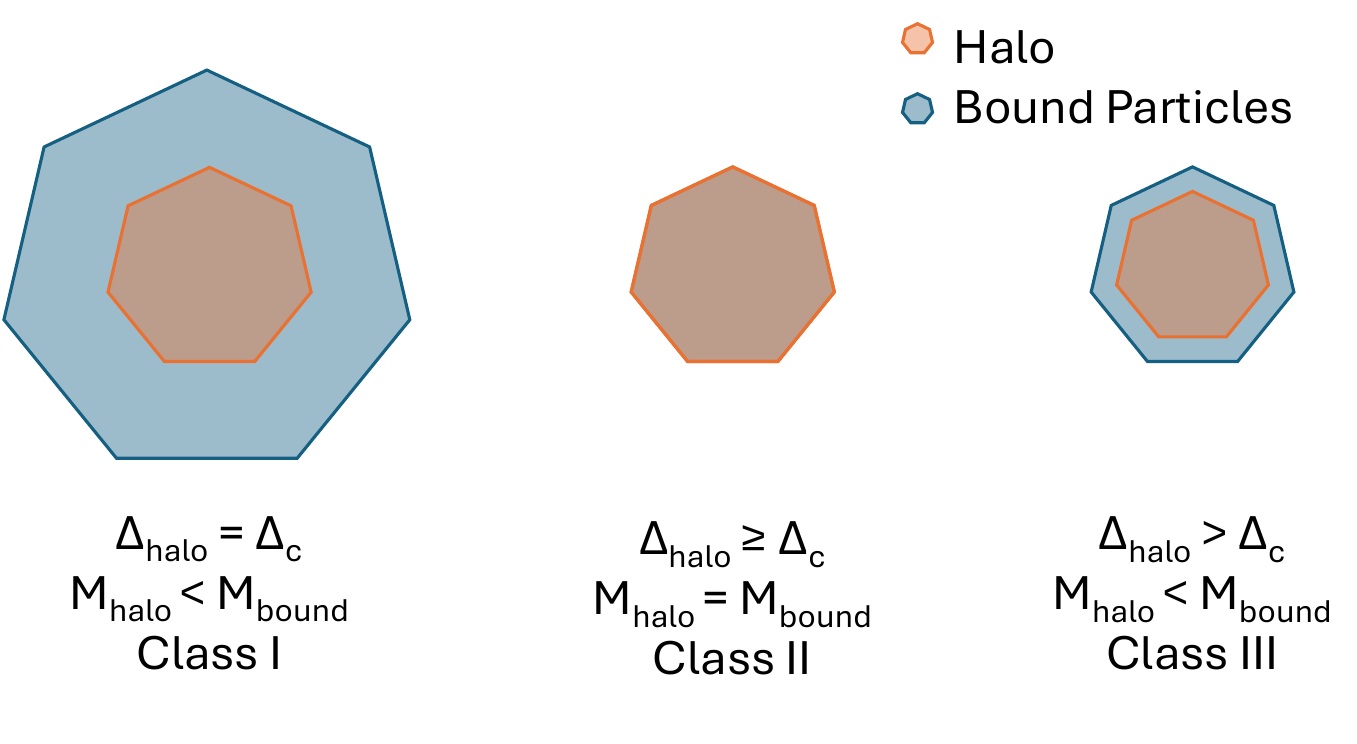}
\caption{Illustration of the three classes of halos described in Sec. \ref{sec: non-sphere} showing the relationship between convex hull halos and the total volume encompassing self-bound particles. Class I halos are classical halos and Class II halos have higher bound overdensities than the virial overdensities.  Class III halos are halos that are recorded at higher overdensity than the virial overdensity target but are also a subset of the bound particles. This occurs as a result of keeping the halo characteristics consistent between timesteps.}
\label{fig: classes}
\end{center}
\end{figure}

In Fig. \ref{fig: comp1} (left) we see the resulting particles seated within their virial radii \citep{1998ApJ...495...80B} (Eq. \ref{eq: BN}) as described in Tab. 1 of \citet{2026ApJ...999...72B}, which were calculated by finding a radius from the halo center that contains an overdensity equal to or similar to the virial overdensity. However, as shown in the figure, the particles are not typically spherically distributed in either small or large halos. Since our bound particle lists are only biased by gravitational energy, we can use them to generate a definition of halo boundedness that completely bypasses the assumption of spherical symmetry.

Based on a target overdensity (see Tab. 1 of \citet{2026ApJ...999...72B}) , our non-spherical halos are defined using the following process. First, a convex hull is defined that encompasses the bound elements of the dark matter sample, and the overdensity of bound particles within the hull is measured. If that value is greater than the target overdensity, then the bound elements of the halo are assumed to be denser than Virial Theorem implies, and no further refinement of the hull is attempted. Otherwise, if the overdensity is lower than the target, each particle is weighted by orbital energy times the inverse square of radius ($E/r^2$) from the center of energy of the halo. All particles with $E/r^2 >{\rm cut\ off_i}$ are discarded from the halo and the overdensity of the remaining particles is calculated. Using this weighting, the final cut-off threshold is solved using gradient descent over up to 50 iterative steps or until convergence to within 5 $\rho/\rho_c$ of the target overdensity is reached. 50 iterations was four times as large as needed for any individual halo to converge in our test sample of about 1,000 halos.

\begin{figure*}
\begin{center}
\includegraphics[width=\linewidth]{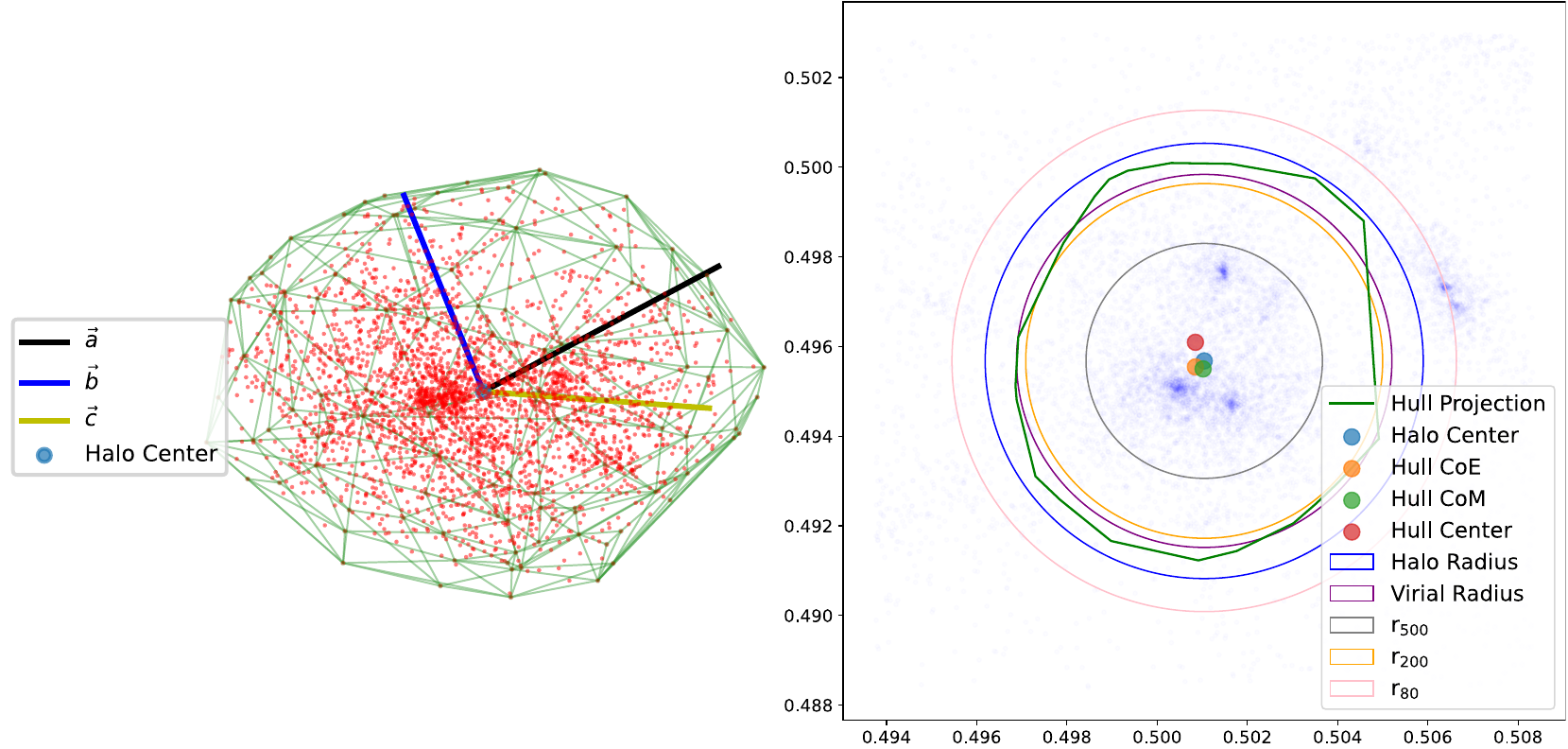}
\caption{Left: The edges and vertices of the convex hull (green) enclosing the bound particles (red) that meet the virial overdensity threshold plotted with the semi-major (black), semi-minor (blue), and intermediate axis (yellow). Right: A 2-D projection of the same hull (green) plotted with a projection of the particles (blue) along with various definitions of center: the center of energy of all the bound particles used as the halo center, the center of energy of the particles included in the convex hull (Hull CoE), the center of mass of the particles included in the hull (Hull CoM), and the average position of the vertices of the hull (Hull Center). Also included are radius measures based on spherical overdensity ($r_{80c},r_{200c},r_{500c}$), the maximum half-width from the halo center to the farthest convex hull vertex (Halo Radius), and the spherical overdensity that corresponds to the overdensity of the convex hull (virial Radius). This convex hull represents a system of halos with a potential that is offset from any of the three prominent self-bound density peaks that are tracked as separately as subhalos.}
\label{fig: hull1}
\end{center}
\end{figure*}

Following this procedure, our halos are only indirectly defined by radius and overdensity and are directly defined by thresholds in orbital energy and distance. This combined energy and distance weighting means that outlying particles of equal boundedness are disfavored, and equidistant particles that are less bound are also disfavored, resulting in a compact structure of tightly bound particles. As shown in Fig. \ref{fig: hull1} (left), the hull boundaries are also bound particle positions, and the hull can contain an unlimited number of vertices and faces. Because all particles evolve dynamically, the membership and shape of the hull are likewise dynamic. An alternative to our use of convex hulls is the use of a non-convex bounding volume such as an alpha shape \citep{1983ITIT...29..551E}. We prefer a convex to a non-convex shape to ease our interpretation of halo volumes and merging as well as to speed our calculations, but we note that method like an alpha shape may be more suited to match the shape of heavily tidally disrupted halos or streams.

Importantly, neither are every particle nor every bound particle inside the convex hull necessarily counted as part of the dense mass of the halo. This is useful especially in the case of mergers where the membership of the hull naturally evolves to encompasses and include infalling material. Thus, our methodology does not presume spherical halos in any step of the finding process, and highly non-spherical halo solutions appear in our results. The spherical version of {\sc Haskap Pie} was designed to track infalling halos and subhalos longer and more consistently than other methods and remains intact as an optional mode. This non-spherical energy-based method is not directly comparable using the same metrics since the foundational definition of a halo has been significantly altered. Therefore, this work is not focused on sub-halo tracking efficacy or comparisons to other halo-finders, which were covered in the antecedent paper, but on exploring the impact of changing the definition of halos and the morphology of the dark matter distributions found with {\sc Haskap Pie}. We emphasize that only dark matter is used to construct all the relationships explored here, including overdensity, but that {\sc Haskap Pie} can include all particles in its calculations.

Our method reports the virial radius as well as other overdensity radii in addition to the enclosing convex hull but internally tracks the halo using the tightly enclosed hull and its overdensity. In many cases, this leads to better tracking of halo objects in complex potentials. In Fig. \ref{fig: comp1} (right), where all the parameters are the same except that the halos are tracked with convex hulls rather than virial radius, we can see that some smaller isolated halos and subhalos are made present.

Our convex hull-finding procedure produces three classes of halos as illustrated in Fig. \ref{fig: classes}. \textbf{Class I} are classic halos wherein the enclosed density of the halo decreases with radius, and the virial overdensity is a subset of the full set of bound particles. For these halos, lower overdensity at larger radii may also be defined, such as $r_{50c}$, and assumptions based on NFW profiles generally hold.

\textbf{Class II} are halos that consist of all the bound particles in a cluster, have a classic inverse relationship between radius and enclosed density, but have a higher overdensity than the Virial Theorem target. Any attempt to define a virial radius for these halos necessarily includes unbound particles and could include mass from another, larger halo, which would disrupt the shape of the density profile. Several halo-finding methods would report halos of this nature as much larger than their actual size by searching for a lower density at a higher radius. This can lead to an inability to continue track the halo if it is embedded in a complex system or cluster of halos and sub-halos.

\textbf{Class III} are halos that have higher overdensity than the virial overdensity but do not encompass all the bound particles. This can occur for both methodological and dynamical reasons. Methodologically, the {\sc Haskap Pie} steps the target overdensity towards the virial overdensity so there are instances when the overdensity of the halo is less than the minimum overdensity of the bound particles. However, for this to occur, the encompassing set of self-bound particles need to be changing more rapidly than the corresponding stepping of the target overdensity. This can occur in complex gravitation potentials such as major mergers and for subhalos moving in and out of the denser parts of a larger halo. Also when the gravitational potential is complex, solutions for a halo are less stable and degenerate, which may lead to larger changes during our progenitor and descendant halo selection process in halo-tracking.



Defining halos based only on an open search of bound particles are intrinsically non-deterministic, as both potential energy and kinetic energy are functions of the set of particles included, and since the particle sampling procedure uses a random assignment. These results converge in simple potential wells, but solutions for more complex structures may change in each iteration. Thus, while results can be reproduced, using different random seeds returns slightly different results. As discussed in \citet{2026ApJ...999...72B}, this is mitigated by producing a large number of candidate halos and using a cost function to prune them until we create a best fiducial tree. On average, several realizations of each halo are available to be pruned to the best solution according to a cost function. In Class III halos, multiple solutions are typically reported in the sample, and a choice must be made about the priority given to the solution with a smaller or larger volume. We have chosen to leave the decision to the same cost function, which penalizes both inconsistent overdensities and radii across timesteps, and allows our solutions to bias towards consistency. As a consequence of our decision to not hard-code the choice into our halo-finding, it is possible for halos to switch between these solutions during high mass ratio mergers.

As shown in Fig. \ref{fig: comp1}, there are cases where the spherical solution reports more halos in an area as well as cases where the non-spherical model reports more halos. Occasionally, when a halo is highly non-spherical, the spherical finder is likely to report the halo as two or more individual halos instead of one. This behavior can potentially produce a more physical description of a halo, such as when the difference is solely due to a discrepancy between the physical shape of the halo and the assumption of a sphere. It can also lead to the loss of merger remnants after major mergers due to the convergence between the two of their potential wells and the energy-based definition of their halo centers, which can happen more quickly when defining a halo based on orbital energy thresholds from the outside in rather than from the inside out based on radii.

\subsection{Halo Definitions}

With this method of finding and tracking halos, a few halo descriptors such as the halo position, halo radius, overdensity, and halo mass lack a consistent definition, so we need to pay special consideration to how they are being used in this context and how they manifest in our results.

\subsubsection{Halo Centers}

Since our halos are non-spherical, several potential definitions of a ``halo center’’ are valid depending on the intended analysis and which particles are counted within a halo. The weighted average position of particles of type $x$, $\vec{r_x}$, over scalar $S$, $\bar{\vec{r}}_S$, can be defined as

\begin{equation}
\bar{\vec{r}}_S = \frac{\sum \limits^{x} S\vec{r_x} }{\sum \limits^{x} S},
\end{equation}

for various choices of $S$ and $x$. If $x$ is all particles that are gravitationally bound to the halo and $S$ is the total energy of those particles, we recover the center of energy. We find that the center of energy is more stable between timesteps than if we used $S$ to be the particle mass to define a center of mass. Therefore, we refer to the center of energy as the ``halo center" and use different descriptors for other measures of the center. The halo finder uses this halo center to track and predict the movement of the halo as well as to determine likely progenitor and descendant halos to build a halo tree.

A common practice in SO and FoF halo-finders is to base or build the center of halos around density peaks.  This presumes that the center of a virialized dark matter particles and its densest region are coincident, and if energy solving is used, it also presumes that the density peak is also the bottom of the potential well. However, both their center of mass or energy can differ significantly from the location of the density peak. This is common in systems that have multiple density peaks due to a population of subhalos or inflalling mergers. Additionally, because multiple density peaks can exist within a complex system of merging halos and subhalos, discontinuities in halo positions can form over time as the central densities evolve. A density peak definition also tends to be sensitive to smoothing lengths and simulation resolution, which act to limit the height of density peaks. Conversely, the center of energy strongly weights the bottom of the potential well, which is less prone to discontinuities. Fig. \ref{fig: hull1} (right) shows the density of the underlying particle distribution (blue markers) in our test simulation as compared to the halo center showing three clear density peaks within $r_{500c}$ and a halo center between them. Each of the density peaks comes from a major merger and will take an extended time to settle into the potential well and so neither of the peaks is well suited to act as the center of the larger halo. Each of the density peaks is captured, however, as a separate halo (Fig. \ref{fig: comp1}, shown as subhalos of the central halo, which is colored in green). Therefore, encompassing halo is a distinct structure that is unconnected to a density peak. This behavior often appears in our results around major or multiple mergers since our energy-solving method will capture cluster-scale structures regardless of the complexity of the particle distribution.

In addition to reporting the center of energy for all the bound particles, we also report centers based on $x$ being the particles that meet the total energy threshold that corresponds to a target overdensity used to find the convex hull and $S$ taking the value of either total orbital energy or particle mass. This returns a center of energy and a center of mass of the particles used to define our halos. Finally, the geometric center of the convex hull can be determined from the hull vertices, which are reported in the finder output. Because these centers might be determined using a subset of the full list of bound particles (Class I described above), it is possible for the halo center to be offset from the other centers as shown in Fig. \ref{fig: hull1} (right).

\subsubsection{Overdensity and Halo Radius}

\label{sec: overmass}

\begin{figure}
\begin{center}
\includegraphics[width=\linewidth]{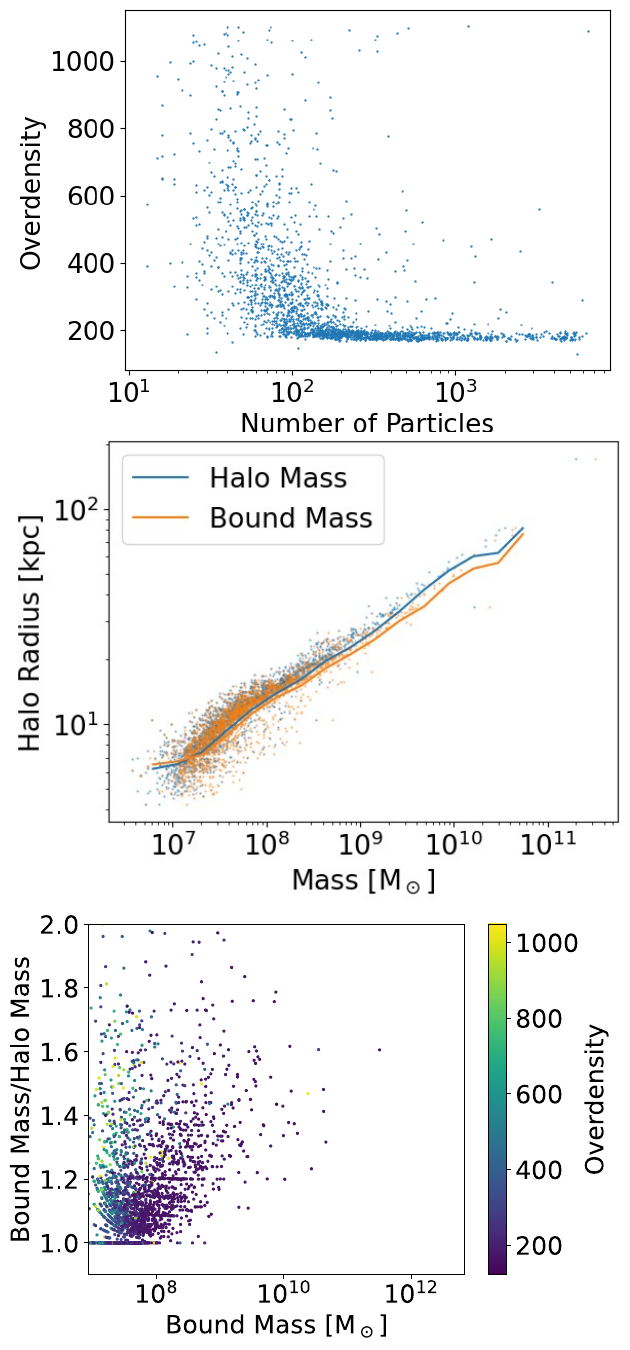}
\caption{Results from the last timestep of the \texttt{CosmoRun}-ART simulation ($z\sim0$) showing quantities related to overdensity, radius, and bounded mass. Each scatter point is a single halo. Top: The overdensity versus particle sample count relationship showing an asymptotic trend at low counts. Middle: The relationship between halo radius and mass showing the total bound mass (orange) and the halo mass within the convex hull. Bottom: The ratio between the bound mass and the halo mass colored by overdensity. Three classes of halos appear in the plot as described in Sec. \ref{sec: non-sphere}.}
\label{fig: overrad}
\end{center}

\end{figure}

\begin{figure*}
\begin{center}
\includegraphics[width=.7\linewidth]{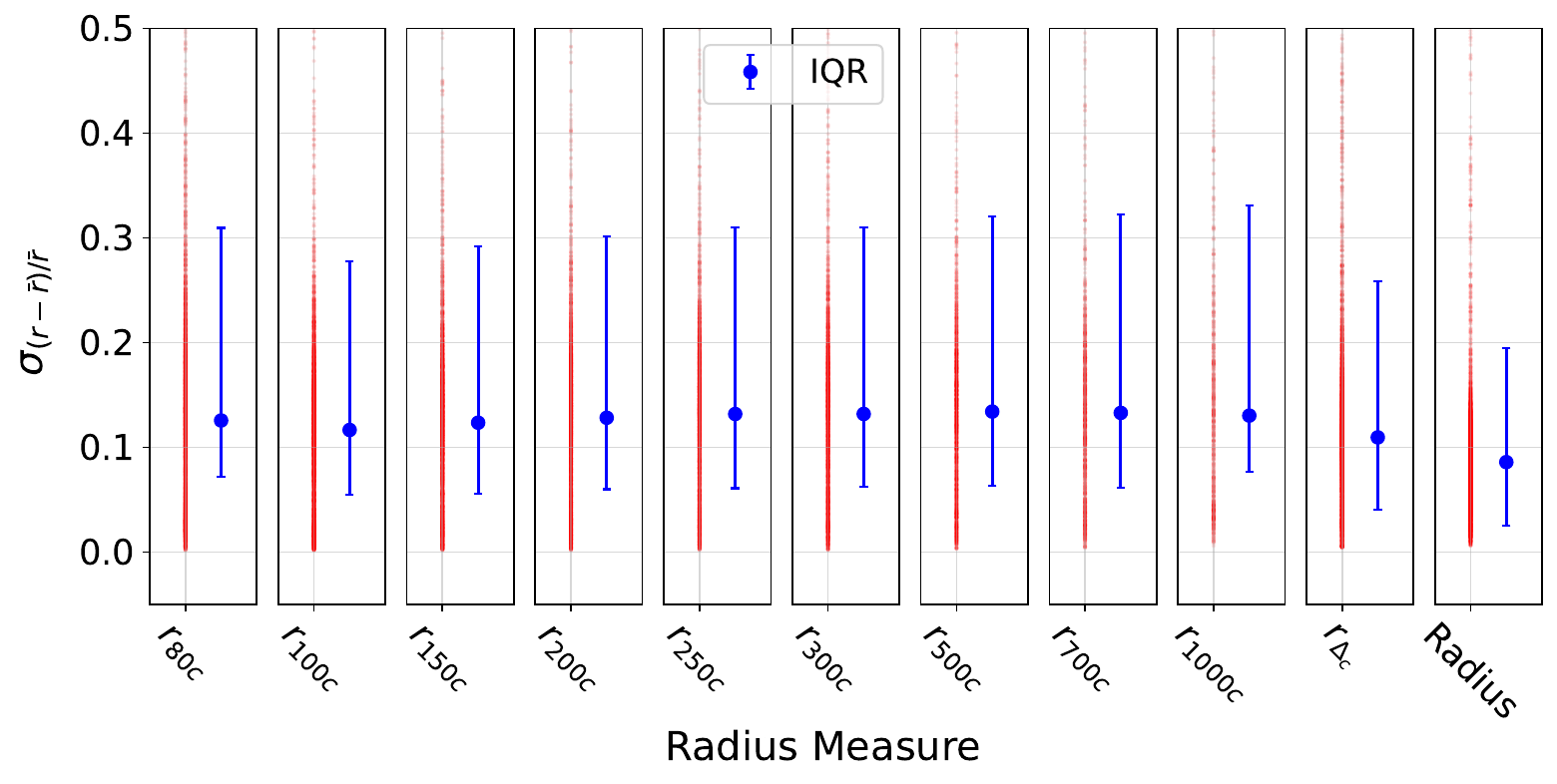}
\caption{Standard deviation of scaled residuals $(r-\bar{r})/\bar{r}$ from a smooth trend, $\bar{r}$, for each measure of radius reported. The smooth trend is generated using a Savgol filter with a window size of 10 and a polynomial order of 3. Each red scatter point represents the standard deviation of all residuals for one halo. The interquartile range and median values of all halos tracks in the \texttt{CosmoRun}-ART simulation are given in blue. The halo radius based on the maximum distance from a hull vertex to the halo center is the most stable measure.}
\label{fig: radcompare}
\end{center}
\end{figure*}

Our hull-finding procedure biases the overdensity of halos towards the virial overdensity, but typically this is only well-defined for halos with sufficient mass and number of particles. Smaller halos tend to fall into Class II, and their bound particle population is more compact. In the first {\sc Haskap Pie} paper, the spherical halo-finding results showed that the number density of halos was suppressed for halos with fewer than around 500 particles. We found halo overdensity to be strongly related to the number of particles that comprise a halo. Fig. \ref{fig: overrad} (top) shows the overdensity-particle number relationship for the final timestep of the \textit{AGORA} \texttt{CosmoRun}-ART simulation. Here, the particle counts are after we've applied our sampling procedure on particle-dense regions. We find that the overdensity of the bound halos grows asymptotically at lower particle number, essentially prohibiting solutions for halos with a few tens of particles. 

This relationship implies that small halos must be compact to be self-bound. This follows from the definition of orbital energy. For kinetic energy to be equivalent to gravitational potential energy, the square of the velocity scales with 1/r. However, the initial dispersion of dark matter velocities is set by the cosmological parameters, so weak potential wells must be dense and deep to capture dark matter. At low particle counts, the depth of the potential well is limited by resolution, and eventually, self-gravitation cannot be determined.

Virial overdensity is historically tied to the premise of spherical non-linear collapse occurring when the overdensity reaches a critical value. In that formalism, the overdensity radius can be described as the radius that encloses a spherical volume that has a given overdensity. For example $r_{200c}$ would be the radius corresponding to an overdensity $\Delta_c = \rho/\rho_c = 200$, which is often used as an approximation of the virial value. However, the critical overdensity for collapse is also a function of scale and sensitive to resolution. Therefore, we accept high overdensity, low mass objects as ``halos’’ since their collapse has been confirmed gravitationally but make no attempt to define them to lower overdensities. As a result, quantities like $r_{200c}$ are not present for all halos. All halos with solved potentials do have a reported overdensity and an overdensity radius, which are defined as the overdensity of the convex hull  and the radius from the halo center that encloses the same overdensity respectively. In Fig. \ref{fig: overrad} (center), we show the relationship between radius and halo mass.

We also define a ``halo radius’’ as the maximum distance from the halo center to a vertex of the enclosing convex hull. This measure is not a true radius, but it does describe the radius of a circumscribing sphere drawn around the hull. As shown in Fig. \ref{fig: radcompare}, this halo radius is more stable between timesteps than radii based on spherical overdensity. Since not all halos have a high peak density due to particle resolutions as well as dynamics, fewer halos have a well-defined radius at higher overdensities.

\subsubsection{Halo Mass and Bound Mass}

The quantity we report as ``halo mass'' is the mass of the particles that are under the $E/r^2$ threshold that corresponds to a target overdensity. We additionally report the mass of bound particles at several spherical overdensity radii, which are roughly comparable to halo and virial masses reported by other halo-finders. Our halo mass is either equal to or smaller than the total mass of all bound particles, which we report as the ``bound mass’'.

Fig. \ref{fig: overrad} (bottom) shows the relationship between bound mass and halo mass for the test sample. Higher mass halos are more consistently solved as halos of the Class I halos and the bound mass to halo mass ratio increases with bound mass. This implies that the virial mass of halos is a smaller fraction of the bound mass as halo mass increases. At the low mass end, we see two behaviors. First, we see a collection of halos that have equivalent halo and bound masses. These are Class II halos where the bound mass is highly overdense. The second behavior is halos that have bound mass to halo mass ratios greater than one. These are halos that fell into the Class II at a recent timestep but now fall into Class III. Since our forward and backward modeling technique gradually steps the target overdensity towards the virial overdensity, halos will continue to have high overdensities for some time, even if more bound particles are present at lower overdensity. We found after extensive testing that this gradual change in the target overdensity lead to much more consistent halo trees, especially in situations where subhalos enter and exit the potential of larger halos.

In this single-timestep sample, about 18\% of halos have equivalent bound and halo masses and the highest ratio of bound to halo mass is approximately 4.91.

\subsection{Measures of Morphology}

\label{sec: measures}

\begin{figure*}
\begin{center}
\includegraphics[width=\linewidth]{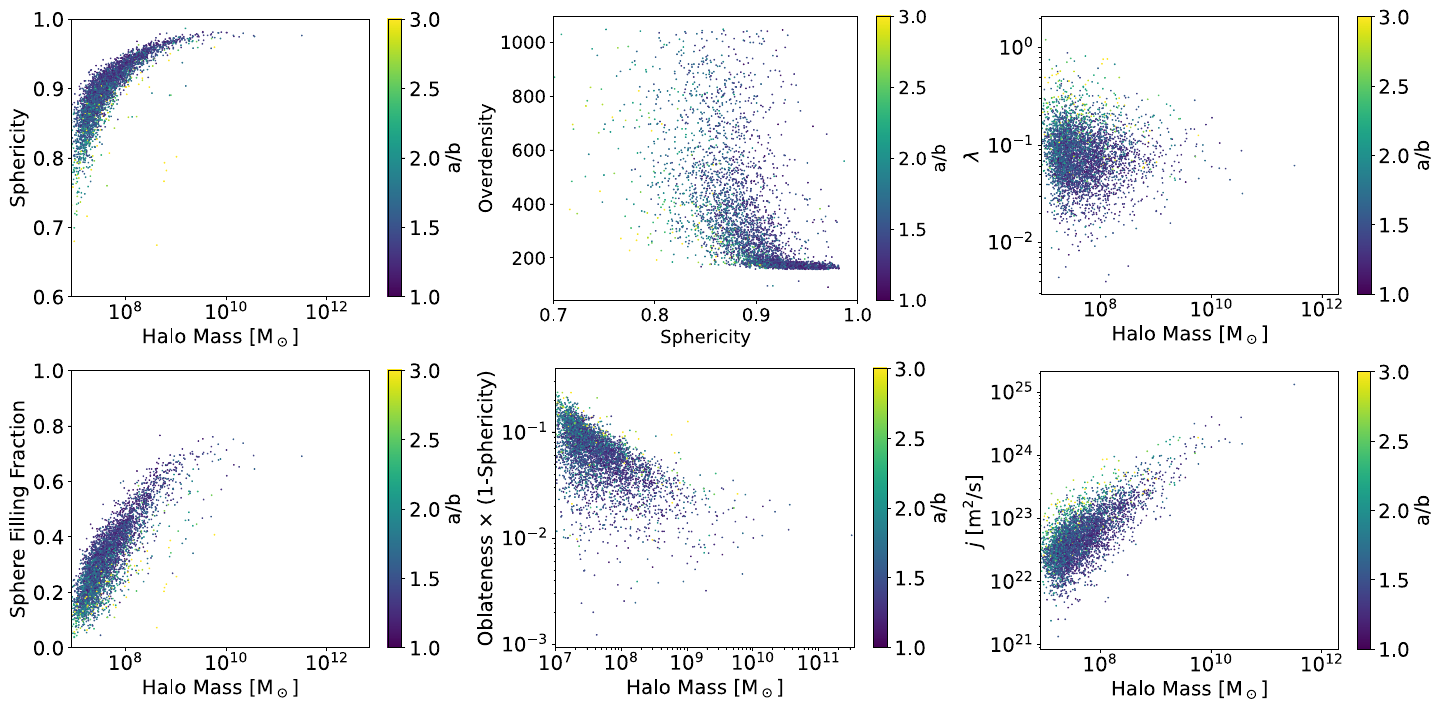}
\caption{Trends in halo morphological features of the last timestep of the \texttt{CosmoRun}-ART simulation, where each scatter point represents a halo all colored by the semi-major axis to semi-minor axis ratio $a/b$. Each of the measures is as described in Sec. \ref{sec: measures}.}
\label{fig: combined}
\end{center}
\end{figure*}

We have adapted, adopted, and produced various measures to study the morphology of our halos. We have consolidated their definitions in this section for reference. Tab. \ref{tab:variables} has also been provided as a concise summary.

\subsubsection{Principal Axes}

Though several other halo-finding methods unbind particles to refine their halo definitions, they are limited in their ability to solve for triaxiality because they do not contain a robust solution for boundedness. However, in post-processing of spherical halos, it is possible to obtain a solution for the semi-major, semi-minor, and intermediate axes by, for example, using the \citet{2006MNRAS.367.1781A} method, which weights particles by the inverse square of their distance from the halo radius.

We find the axes directly by solving the eigenvectors of the covariance of the halo particle positions, which returns the orientation of an orthogonal set of principal axes. For halos that use particle sampling, we recreate the original particle distribution before taking the covariance. The square root of the eigenvalues returns the relative length of the axes, which we divide by the highest value and multiply by the halo radius as well as the number of particles. This is then multiplied to the unit vectors in the direction of the corresponding eigenvector direction, which returns the semi-major axis, $\vec{a}$, the semi-minor axis, $\vec{b}$, and the intermediate axis, $\vec{c}$. These vectors are reported in the halo-finding results and are demonstrated in the example shown in Fig. \ref{fig: hull1} (left). The classical scalar values of these axes ($a,b,c$) can then be determined by taking the Euclidean norm of these vectors. By construction, $a$ is equal to the halo radius. Therefore, unless the semi-major axis is exactly aligned to the largest half-width, $\vec{a}$ will expand a bit beyond the boundary of the halo. The ratio of its length with respect to the other axes is consistent regardless of scaling and can be directly used for measures of morphology.

\begin{table*}[ht]
\centering
\caption{Definition of key variables used to describe halos with {\sc Haskap Pie}.}
\begin{tabular}{c|c|l}
Variable & Symbol & Short Description  \\
\hline
Halo Center & N/A & Center of energy of the bound particles  \\
Center of Hull & N/A & Mean position of all vertices of the convex hull halo \\
Halo Radius & N/A & Longest distance from the halo center to a vertex of the convex hull halo \\
Halo Mass & N/A & Mass of the bound convex hull halo particles \\
Bound Mass & N/A & The total mass of self-bound particles \\
Overdensity Radius & $r_{\Delta_c}$ & Spherical overdensity radius corresponding to the convex hull halo overdensity \\
Spin & $\lambda$ & Total spin of the convex hull halo \\
Sphericity  & $\Psi$ & Ratio of surface area of a sphere to the surface area of the convex hull halo  \\
Sphere Filling Fraction  & $\phi$ & Ratio of the volume of a sphere to the volume of the convex hull  \\
Semi-major Axis & $\vec{a}$ & Longest principal axis of the convex hull halo \\
Semi-minor Axis & $\vec{b}$ & Shortest principal axis of the convex hull halo \\
Specific Angular Momentum & $\vec{j}$ & Angular momentum of particles in the convex hull halo divided by their mass  \\
Spin & $\lambda$ & Total spin of the convex hull halo \\
Oblateness & $O$ & Angle between $\vec{a}$ and $\vec{j}$ scaled by $\pi/2$ (Range: [0,1]) \\
\hline
\end{tabular}

\label{tab:variables}
\end{table*}

This is slightly similar to the \citet{2006MNRAS.367.1781A} method since a 1/r$^2$ weighting was used to determine the extent of the convex hull. However, they are related to particle clustering in linear order rather than quadratic order, which is a key difference. Our axes are not an expression of how the closest particles are weighted but are a measure of the full shape of the overdense region. While we calculate these axes, we note that since our halos can take on highly non-ellipsoidal shapes, $a,b,$ and $c$ do not have a consistent meaning. In our analysis, we focus on the $a/b$ ratio ($s$ in other work) incorporate departures from a spherical shape to combinations of other measures.

\subsubsection{Sphericity and Sphere-filling Fraction}
\label{sec: sphere methods}

To detect how our halos diverge from a spherical morphology, we have adapted two measures. Sphericity is defined as the ratio between the surface area of a sphere of volume $V$ and the surface area, $A$, of an object of the same volume,

\begin{equation}
\Psi = \frac{\pi^{1/3}(6 V)^{2/3}}{A}.
\label{eq: psi}
\end{equation}

This measure is unity if the object is a sphere and less than one for all objects that are not spheres. The number of sides can limit the theoretical maximum value of sphericity. For example, a cube has a sphericity of $\approx 0.806$ and a dodecahedron (regular 12-sided polyhedron) has a sphericity of $\approx 0.911$. This means that the number of particles describing a convex hull can place an upper limit on this measure, but only for small numbers of particles. As shown in Fig. \ref{fig: combined} (left top), which shows the halos in the final timestep of the \textit{AGORA} \texttt{CosmoRun}-ART simulation, there is a clear positive relationship between sphericity and halo mass. In that timestep, the highest value of sphericity is $\approx 0.987$.

Alternatively, we can define a sphere-filling fraction as

\begin{equation}
\phi = \frac{3V}{4\pi r^3},
\label{eq: phi}
\end{equation}

where $r$ is the halo radius. This is a measure of the fraction of a sphere with the same radius as the maximal distance between the halo center and the farthest vertex is filled by the hull volume and emphasizes vertex outliers. As shown in Fig. \ref{fig: combined} (left bottom) for the same timestep, the trend in this measure is more pronounced and less likely to asymptotically approach unity for high sphericity halos.

\subsubsection{Angular Momentum and Spin}

The spin parameter is classically defined as

\begin{equation}
\lambda = \frac{J|E|^{1/2}}{GM^{5/2}},
\label{eq: spin}
\end{equation}

where $J$ is magnitude of the total angular momentum, $E$ is the total orbital energy, and $M$ is the mass of the system. Traditionally, it has been difficult to directly determine the spin parameter of halos using results from other halo finders since the orbital energy, $E$, is typically not directly calculated for each particle. Instead, alternate measures such as the \citet{2001ApJ...555..240B} spin have used assumptions of an NFW profile, and a spherical halo have been developed and heavily used in the literature. In the formalism for that parameter, they motivate the development of their spin parameter by noting that the energy distribution of halos is more complex than the set of assumptions used to calculate it. Studies such as \citet{2017MNRAS.466.1625Z} have examined the difference between the spin parameter and found Eq. \ref{eq: spin} to be more consistent across redshift (Fig. 9 in their paper), however the comparisons were made using SO and FoF halo-finders rather than a fully non-spherical methodology.

Conversely, our halo finding methodology determines the energy of each particle as a fundamental step of our solving process. This allows us to recover the true spin parameter and report $\lambda$ and halo-specific angular momentum, $j = J/M$, with our halo-finding results. These results include the caveat that we take angular momentum from the center of energy instead of the center of mass, since this center seems to be closer to the true center of orbital motion. In Fig. \ref{fig: combined} (right top), we show that the distribution of $\lambda$ is positively correlated $a/b$, and the spread of $\lambda$ in log space is negatively correlated with halo mass. The distribution of $j$ is positively correlated to both halo mass and $a/b$ as shown in Fig. \ref{fig: combined} (right bottom).

\subsubsection{Quantities Derived from Principal Axes}

\begin{figure*}
\begin{center}
\includegraphics[width=1\linewidth]{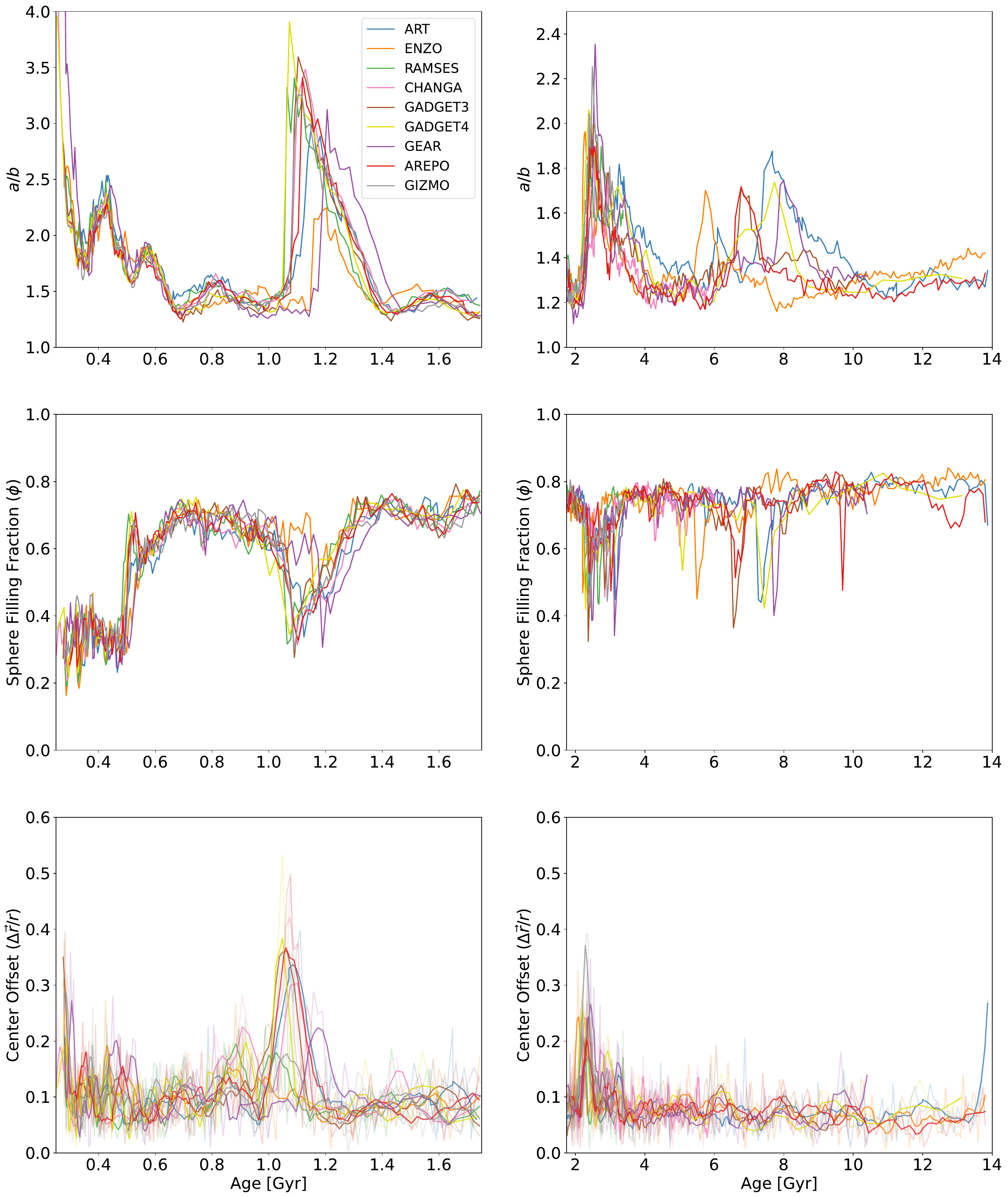}
\caption{The evolution of three shape parameters for the main halo in each of the nine simulations. Left plots show early times, which include a high mass ratio major merger. Right plots show the late-time evolution. Top: $a/b$ ratio showing peaks associated with mergers. Middle: sphere filling fraction (Eq. \ref{eq: phi}) showing a more impulse-like responsiveness to minor mergers. Bottom, the offset between the center of energy of the halo and the geometric center of the halo scaled by the halo radius. In the bottom plots, the full trend is shown as transparent lines and the smoothed trend is shown as opaque lines.}
\label{fig: ab}
\end{center}
\end{figure*}

Oblateness and prolateness can be determined directly from the principal axes. Oblateness is a measure of how normal the angular momentum vector is to the semi-major axis vector or

\begin{equation}
O = \frac{2}{\pi}cos^{-1}\left|\frac{\vec{j}\cdot\vec{a}}{|\vec{j}||\vec{a}|}\right|,
\label{eq: oblate}
\end{equation}

where values greater than 1/2 indicate an oblate object with the semi-major axis that is more normal to the angular momentum axis and values less than 1/2 indicate a prolate object. In the example final \texttt{CosmoRun}-ART timestep ($z=0$), we find that $\sim$77.2\% of the halos are oblate and the remaining fraction are prolate. We further incorporate the sphericity of the halo in order to emphasize the shape of the halo by taking $O(1-\Psi)$. High values of this measure indicate an oblate object with low sphericity. The additional factor of $1-\Psi$ eliminates the singularity of the dot product between $\vec{j}$ and $\vec{a}$ for spherical objects. As shown in Fig. \ref{fig: combined} (center bottom), this quantity is typically higher for smaller halos since sphericity is highly correlated to halo mass.

This is distinct from the \citet{1988ApJ...327..507F} definition used in subsequent work, which categorized shape parameters of halos by the ratio of the three axes. Therein, they define a halo as oblate if $c/a < b/c$, prolate if  $c/a > b/c$, and triaxial if $c/a \sim b/c$. Thus, in that definition, a halo is prolate if the intermediate and semi-minor axis are close in magnitude and prolate if they are not.
Because our halos are neither assumed to be spherical nor ellipsoidal in their construction, it can be beneficial to define their shapes such that they respond to departures from symmetry through sphericity in this way.

\section{Results}

\label{sec: results}

The \textit{AGORA} \texttt{CosmoRun} simulations have zoom-in regions focused on the evolution of a Milky Way-sized galaxy. Therefore, we pay special attention to the evolution of that galaxy’s dark matter halo across cosmic time and between simulations in Sec. \ref{sec: mainhalo}. Additionally, we report morphological characteristics for the full sample of halos caught inside the zoom-in region in Sec. \ref{sec: allhalo}. \textit{AGORA} Paper I \citep{2014ApJS..210...14K} already showed that differences in the treatment of baryons between the codes do not strongly affect dark matter dynamics and likely do not play a key role in shaping halo morphology. Prior studies of these data have characterized the evolution of the main halo and compared the merger trees. This showed a timing discrepancy for the first major merger and subsequent major mergers as examined in \citep{2024ApJ...968..125R}. Outside of this, the codes roughly agreed on broad morphological trends.  Therefore, we focus on establishing how the codes agree and differ in morphology due in the context of dynamical dispersions in the evolution of dark matter.

\subsection{Main Halo}

\label{sec: mainhalo}

Some of the timing discrepancies between codes that occurred in the halo tree of the first major merger characterized in previous works could have originated in a halo-finder’s definition of a ``merger.'' As shown in \citet{2026ApJ...999...72B}, the time a halo-finder loses track of a merging halo does not converge and several works have been able to extend the tracking of infalling halos. Therefore, a merger should not be defined by when a halo-finder loses track of a progenitor. An alternate definition of a merger is when the two virial radii begin to overlap. However, while this metric is more stable, it is premised on an assumption of spherical symmetry precisely when halos are least spherical. With our non-spherical results, we can directly study the impact of these mergers on the shape of the halo as well as measure how long disruptions to a spherical halo persist, allowing us to alternatively define major mergers through particle dynamics.

\subsubsection{Principal Axes}

\begin{figure}
\begin{center}
\includegraphics[width=0.9\linewidth]{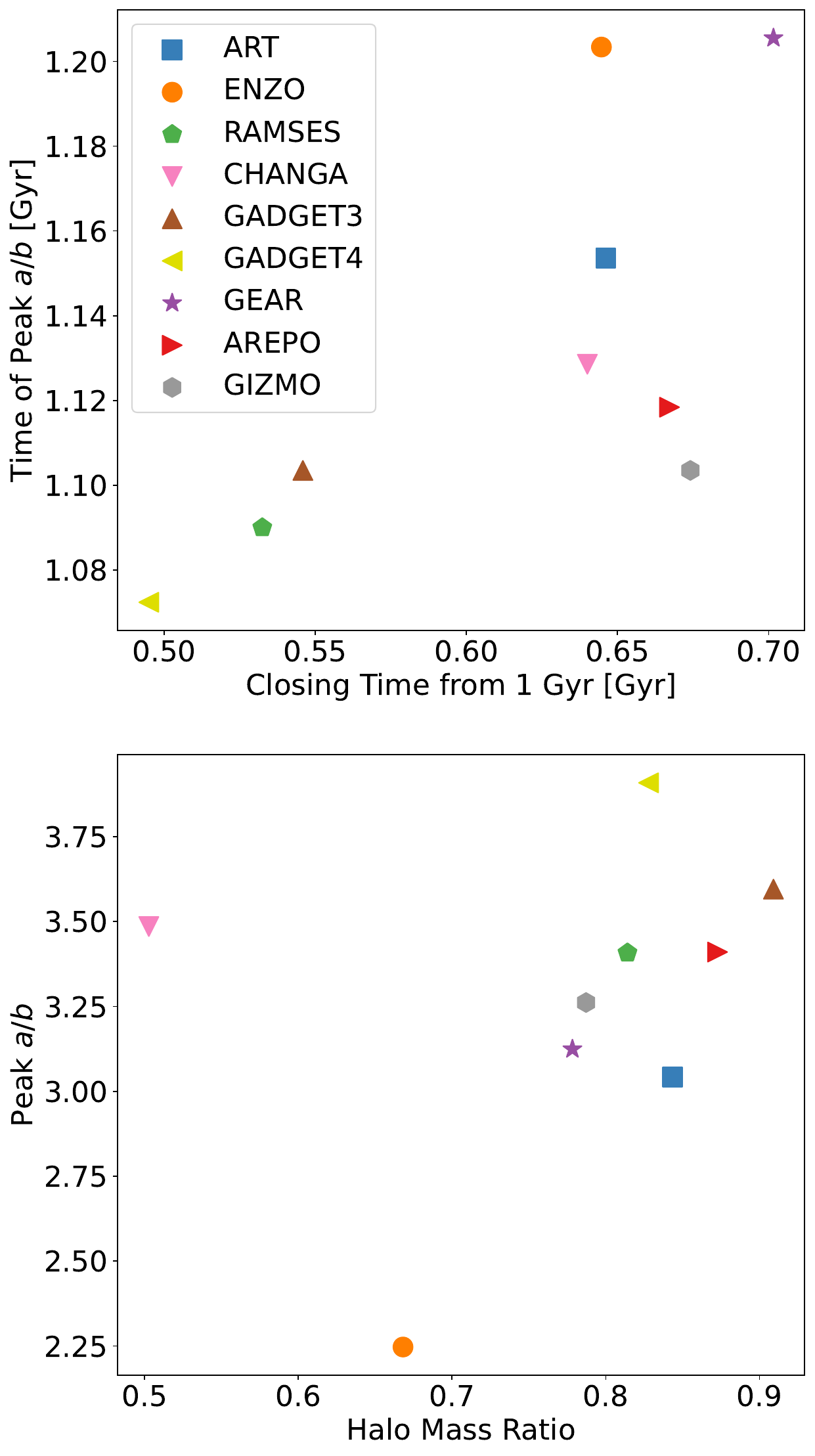}
\caption{Trends connecting the $a/b$ ratio of the high mass major merger at $\sim$1.1 Gyr to the state of the infalling halo at 1 Gyr. Top: the time of the peak in $a/b$ versus the closing time (distance divided by radial closing velocity) showing a roughly positive correlation. Bottom: The peak value of $a/b$ versus the merger mass ratio also shows a positive correlation with CHANGA’s realization as an outlier.}
\label{fig: abtrend}
\end{center}
\end{figure}

\begin{table*}[ht]
\centering
\caption{Table of quantities used to produce Fig. \ref{fig: abtrend} and the cluster ratio, which is defined and explored in the appendix in the context of dense collections of halos simultaneously merging. Distance, closing time, and mass ratio are taken at 1 Gyr. The outlier mass ratio value for CHANGA is explored in the Appendix.}
\begin{tabular}{c|c|c|c|c|c|c}
Simulation Code & Distance [kpc] & Closing Time [Myr] & Mass Ratio & Cluster Ratio & Peak $a/b$ & Peak $a/b$ Time [Gyr] \\
\hline
ART & 29.4 & 646 & 0.844 & 1.32 & 3.04 & 1.15 \\
ENZO & 33.0 & 645 & 0.668 & 0.94 & 2.24 & 1.20 \\
RAMSES & 29.5 & 532 & 0.814 & 1.26 & 3.49 & 1.13 \\
CHANGA & 31.5 & 640 & 0.503$^*$ & 1.25 & 3.48 & 1.13 \\
GADGET-3 & 28.5 & 546 & 0.909 & 1.22 & 3.60 & 1.10 \\
GADGET-4 & 28.7 & 495 & 0.828 & 2.38 & 3.91 & 1.07 \\
GEAR & 34.5 & 702 & 0.778 & 1.16 & 3.12 & 1.21\\
AREPO & 31.0 & 667 & 0.872 & 2.41 & 3.41 & 1.11 \\
GIZMO & 30.4 & 674 & 0.787 & 0.97 & 3.26 & 1.10\\
\hline
\end{tabular}

\label{tab:timing}
\end{table*}

In individual timesteps, larger halos are more likely to be more spherical, have lower values of $a/b$ (see Fig. \ref{fig: combined}, top left), and converge to spherical collapse theory and thus are typically captured in the halo finder as Class I halos. The exception to this is high mass ratio mergers, which can distort the morphology significantly. Mergers appear clearly as large spikes in the $a/b$ ratio of the main halo as shown in Fig. \ref{fig: ab}. The halo-finder tracks the main halo backwards in time and there is rough agreement between codes as to when the $a/b$ ratio starts to grow ($\sim$1.4 Gyr).

The peak in $a/b$ represents the highest distortion that occurs before both halos are regarded as fully distinct self-bound structures. There is some disagreement between simulations as to when the $a/b$ ratio peaks and falls, which occurs anywhere from $\sim$1.05 Gyr to $\sim$1.22 Gyr.  Generally, the longer the two halos are regarded as a single structure, the larger the peak. There is also some correlation between the magnitude of the peak and how early in Cosmic Time the peak occurs for the same reason.

The differences in timing and peak magnitude of the spike may be related to the halo-finder solving method rather than a physical process, so we investigated whether there were dynamical indicators of when a peak might occur that preceded the merger. Specifically, we examined the mass ratio, distance, and closing time of the two halos at 1000 Myr across simulations and compared them to the magnitude and time of the peak of $a/b$. As shown in Tab. \ref{tab:timing}, the closing time, which is defined as the distance divided by the projection of the relative velocity between the two halos onto the line connecting them, was correlated to the time of the peak of $a/b$ (see Fig. \ref{fig: abtrend} top). Furthermore, GADGET4’s version of the merger is a clear outlier, happening earlier than the others due to the halos having a much lower closing time to each other by 1000 Myr, which cannot be explained except as an intrinsic difference in halo positions and velocities.

With the exception of the halo in CHANGA, the merger mass ratio was strongly correlated to the $a/b$ peak (see Fig. \ref{fig: abtrend} bottom). This follows from an argument that a major merger can deform the potential well more than a more minor merger. Given the connection between the $a/b$ ratio and dynamical processes, there is the potential to use the timing of the peak and the time when $a/b$ settles to a minimum might as indicators for the onset of a merger and the coalescence of the two halos respectively, regardless of the efficacy of halo-tracking. A deeper discussion of the CHANGA result is provided in the appendix, were we show that it may be due to a complex system of simultaneous major mergers.

We also investigated the other prominent merger events for the main halo. The mass ratio of the largest merger at around 2.5 Gyr is between 0.16-0.20 (at 2 Gyr) for all simulations except GADGET4 and GEAR and the peaks of $a/b$ are correspondingly lower than the $\sim$1.1 Gyr merger with values between 1.8 and 2.25. GADGET4’s and GEAR’s merger mass ratio were outliers at 0.55 and 0.28 with $a/b$ peaks of 2.06 and 2.35, respectively. The last strong change in $a/b$ is related to a merger that occurs between 5.8 and 8.5 Gyr, depending on the simulation, indicating that the mergers are by then unsynchronized. In each of the three merger events, GEAR’s and ART’s peaks are among the latest, and ENZO’s peaks move from last to first over time.

Outside of the effect of mergers, there does not seem to be any secular evolution in the $a/b$ ratio of the main halo after 700 Myr in any of the simulations and the halos tend to settle to ratios between about 1.2 and 1.4 between disruptions, at all redshifts.  Settling times are a strong function of redshift as lengthened free-fall and dynamical times at later times also manifest as long settling times. Conversely, merger mass ratios do not have an obvious effect on settling times, even among mergers at similar redshifts.

In summary, there are at least two major discrepancies between the dark matter halos that arise when comparing simulations. First, there are differences in the states of the halos (positions and velocities) that subtly accumulate over time, leading to timing discrepancies. In some cases, this difference exceeds any effect that can be attributed to gravity solving, softening length, or general numerical effects. Second, there are differences in the halo merger mass ratios between simulations, which lead to dynamical and morphological differences in the resulting halos that also accumulate over time.


\subsubsection{Sphere-filling Fraction and Sphericity}

\begin{figure*}
\begin{center}
\includegraphics[width=1\linewidth]{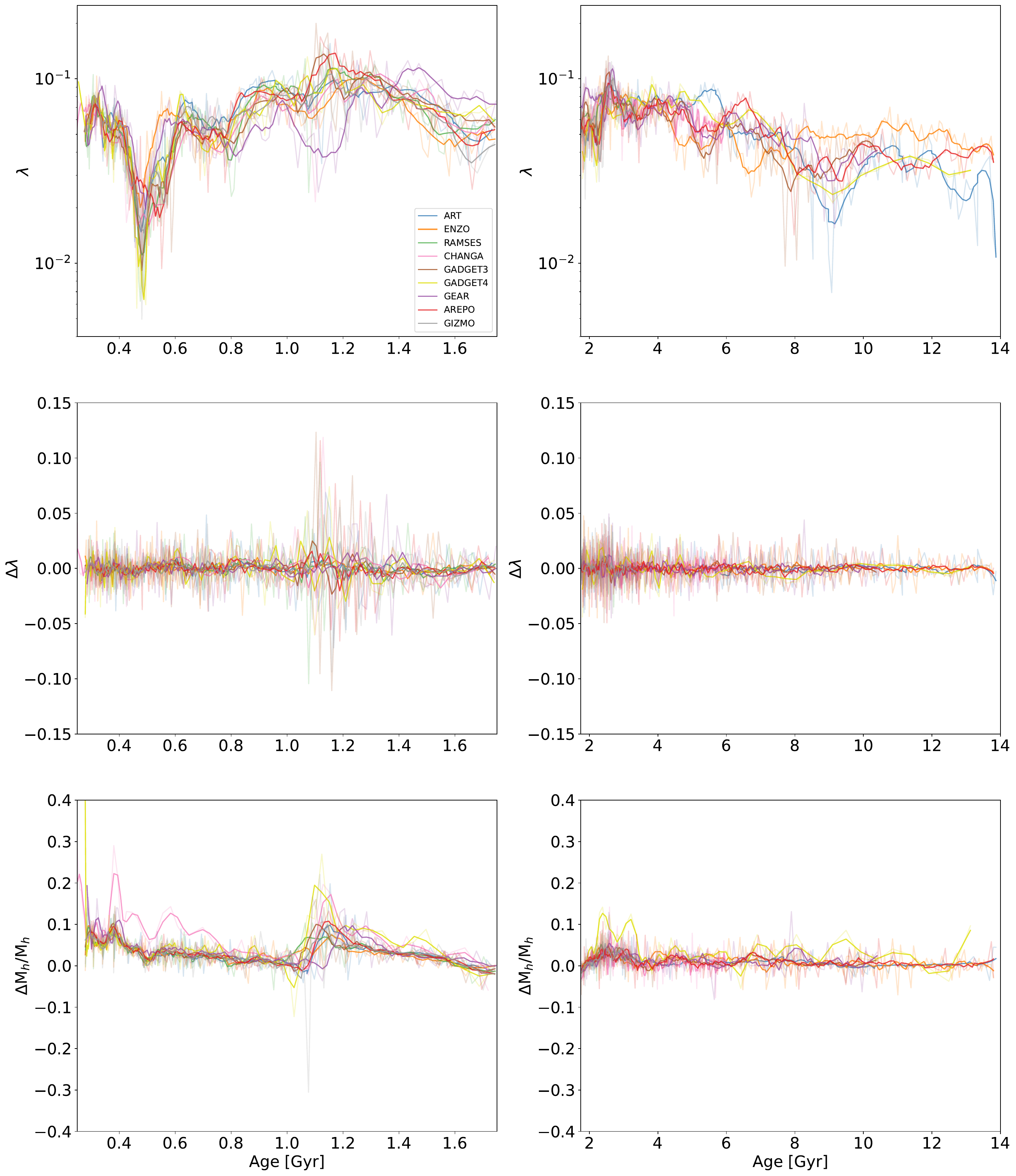}
\caption{Spin ($\lambda$, top), change in spin ($\Delta \lambda$, center), and change in halo mass scaled by halo mass ($\Delta M_h/M_h$, bottom) for the main halo as a function of simulation time. Smoothed values are opaque and the full trend is translucent.}
\label{fig: mainspin}
\end{center}
\end{figure*}

Though Fig. \ref{fig: combined} (bottom left) showed that sphere-filling fraction, $\phi$, was highly correlated with halo mass for the halos in a single low-redshift timestep, we found that $\phi$ of the main halo sharply responds to mergers with high variation. A reasonable expectation is that the process of a merger causes drops in $\phi$ as the $a/b$ spike. However, at low mass and early times, dips in $\phi$ due to mergers are much less prominent than increases in $\phi$ that arise when the higher combined mass after the halo leads allows the halo to settle into a spherical morphology. This effect can be seen at the beginning at $\sim$500 Myr in Fig. \ref{fig: ab} (middle row). As shown in Fig. \ref{fig: combined} (bottom left), the transition typically happens in halos between 10$^8$ and $\sim$5$\times$10$^{10}$ M$_\odot$, which is above the threshold of $\sim$$10^8$  M$_\odot$ ($\sim$350 particles) where overdensity grows asymptotically, and halos switch between Class I and Class II or III. At $\sim$450 Myr, which is prior to the merger that causes the jump and during the $a/b$ peak of a preceding strong merger, the halo’s mass is $\sim8\times10^9 M_\odot$ ($\sim$28,600 particles). This implies that the shape change is more mass-dependent than particle-count dependent.

At higher mass, mergers consistently present as prominent dips in $\phi$ as the morphology of the halo is temporarily disrupted before settling back into a more spherical shape. Because $\phi$ is based only on the shape of the bounding hull and $a/b$ is weighted by particle positions, these two measures can present different signals. This is because halos that merge only directly disrupt the shape of the main halo as they cross the hull boundary, whereas deviations in $a/b$ can persist as long as the center of energy and the density peaks of the two halos are separated. Therefore, changes in $\phi$ are typically sharper and more responsive to low-mass mergers than changes in $a/b$. For example, it is easier to see that the event between 2 and 4 Gyr is actually two high mass ratio mergers in relatively short succession from the double dips. However, the magnitude of the dips in $\phi$ are relatively similar between mergers and so this feature is not as clear a signal of the merger mass ratio as $a/b$.

Outside of mergers at later times, the sphere-filling fraction continues to gradually trend upward. As noted in Sec. \ref{sec: sphere methods}, due to its definition, the sphere filling fraction can only approach unity if the center of the hull and the center of energy are the same and any offset will cap $\phi$ at a lower value. As shown in Fig. \ref{fig: hull1} (right), the hull center can be offset from the center of mass and energy can be offset in a direction that does not tend towards a local density peak.  We find that filamentary structure (which forms a ``v’’ shape  to the top right and to the left in those figures) can pull the hull center towards it as the edges of the hull, and thus the hull center, are more sensitive to smaller density changes at the hull boundary. In \citet{2026ApJ...999...72B} (Figure 12), the bias the filamentary structure that results in the $\phi$ cap for the \texttt{CosmoRun} main halos can be seen in each of the simulations through the clustering of merger infall points.

As shown in Fig. \ref{fig: ab} (bottom row),  we calculated the distance between the center of mass of the hull and its geometric center scaled by the halo radius, $\Delta \vec{r}/r$. Due to its sensitivity to smaller changes in density, this measure had much more variation than $a/b$ between timesteps. To determine a secular trend, we used a third-degree polynomial Savgol filter. For the filter window, we took the number of timesteps between 1 Gyr and 2 Gyr in each simulation and divided that period into five groups (5-13 timesteps per $\sim$200 Myr for this period). We found that the gradual increase in $\phi$ indeed corresponded to a downward trend in $\Delta \vec{r}/r$. As a merger signal, this measure was less sensitive to minor mergers than $\phi$ or $a/b$ and thus better at differentiating between major and minor mergers than $\phi$.

We also explored the evolution of sphericity, $\Psi$, for the main halo across simulations, which generally responded similarly to $\phi$, except the low mass mergers did not appear as strong signals.

\subsubsection{Mass and Spin}

The bound particles in the convex hull used to calculate the halo mass only represent a subset of the bound particles for Class I halo results. To test the variation in the number of particles used to describe a halo, we again used a third-degree polynomial Savgol filter with the same window size to find the secular change, $\Delta\bar{m}$, in halo mass for the main halo. The evolution of $\Delta \bar{m}/\bar{m}$ (dashed line) and  $\Delta m/m$ (solid line) is shown in Fig. \ref{fig: mainspin} (bottom panels). We then calculated the standard deviation of $(\Delta m-\Delta \bar{m})/\bar{m}$ for each mass, $\sigma_{m}$, for the period between 500 and 1750 Myr, which was available for all nine simulations.

Except for GIZMO, which has a large mass adjustment during the major merger, we found that $\sigma_{m}$ was lower for halo mass (0.0175-0.0431) than bound mass (0.0305-0.0526) as the more compact definition of the halo at higher overdensity was found to be more stable. Also, except for GIZMO, larger values in the range corresponded to larger intervals between timesteps in the simulation. In all cases except ENZO, $\sigma_{m}$ for halo mass was less than for $m_{300c}$ (0.0218-1.13) even though $m_{300c}$ was a higher density than the halo density at all timesteps. This is due to an incongruence between spherical overdensity and our definition of a halo that leads to higher variation for measures that assume spherical symmetry. Therefore, our halo quantities like spin and angular momentum are based on the particles used to define our halo mass instead of the bound mass or other definitions for consistency.

Peaks in $a/b$ coincide with large changes to the volume of the halo. Spin peaks and troughs occur during merger events but are not always coincident with $a/b$ peaks, implying that large spin changes are not strictly stemming from changes in the halo boundaries or the particles included. During a merger, the halos may spin down (reduction in $\lambda$), spin up (increase in $\lambda$), or both. The highest peaks in $\lambda$ during the merger at $\sim$1.1 Gyr are GADGET3, AREPO, RAMSES, and ART. In order of minima, the most prominent spin downs were CHANGA, RAMSES, and GEAR. Since a spin-down implies that the halos have opposing angular momenta and a high angular momentum difference can disrupt halo morphology during a merger, CHANGA’s low minimum spin is potentially related to the outlier $a/b$ peak of the interaction. Despite low minima, when spin was smoothed with a Savgol filter, CHANGA and RAMSES had spin changes that were in line with the mean trends of the other codes. The only code to show a prominent secular reduction in spin during the merger was GEAR (Fig. \ref{fig: mainspin}, top left at 1.1 Gyr).

Spin changes during other mergers seem to have better agreement between codes, but some mergers do not induce a significant change in spin. Over the long term, the spin of the main halo evolves most rapidly during the first $\sim$ 1.25 Gyr of the simulations, when it peaks and then trends gradually towards zero. This implies a limited volume of correlated spins about our main halo that is washed out at large scales and higher halo masses.

\begin{figure*}
\begin{center}
\includegraphics[width=\linewidth]{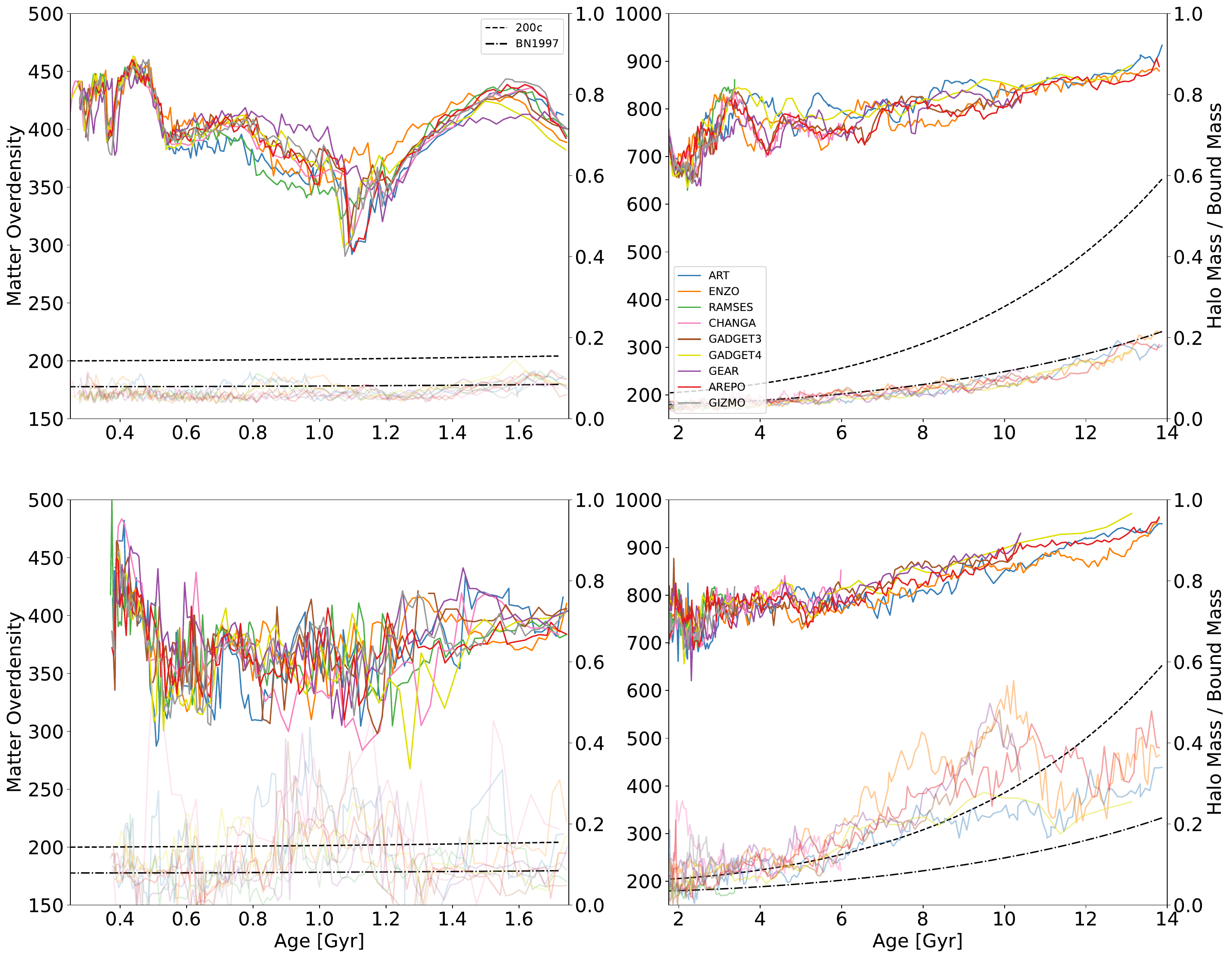}
\caption{Top: halo mass to bound mass ratios (right axis, solid line) for the main halo versus time, as well as halo matter overdensity versus time (left axis, translucent). The main halo tracks the virial overdensity (black dot-dashed line) as designed. Also plotted is 200c as a dashed line.  Bottom: the mean trend of both variables for halos with masses greater than $8 \times 10^8$ and between 0.5 and 3 virial radii from the main halo center, showing how dynamical effects can tend to inflate the recovered matter overdensity and halo mass to bound mass ratio.}
\label{fig: mainover}
\end{center}
\end{figure*}

\subsubsection{Overdensity}

\begin{figure*}
\begin{center}
\includegraphics[width=\linewidth]{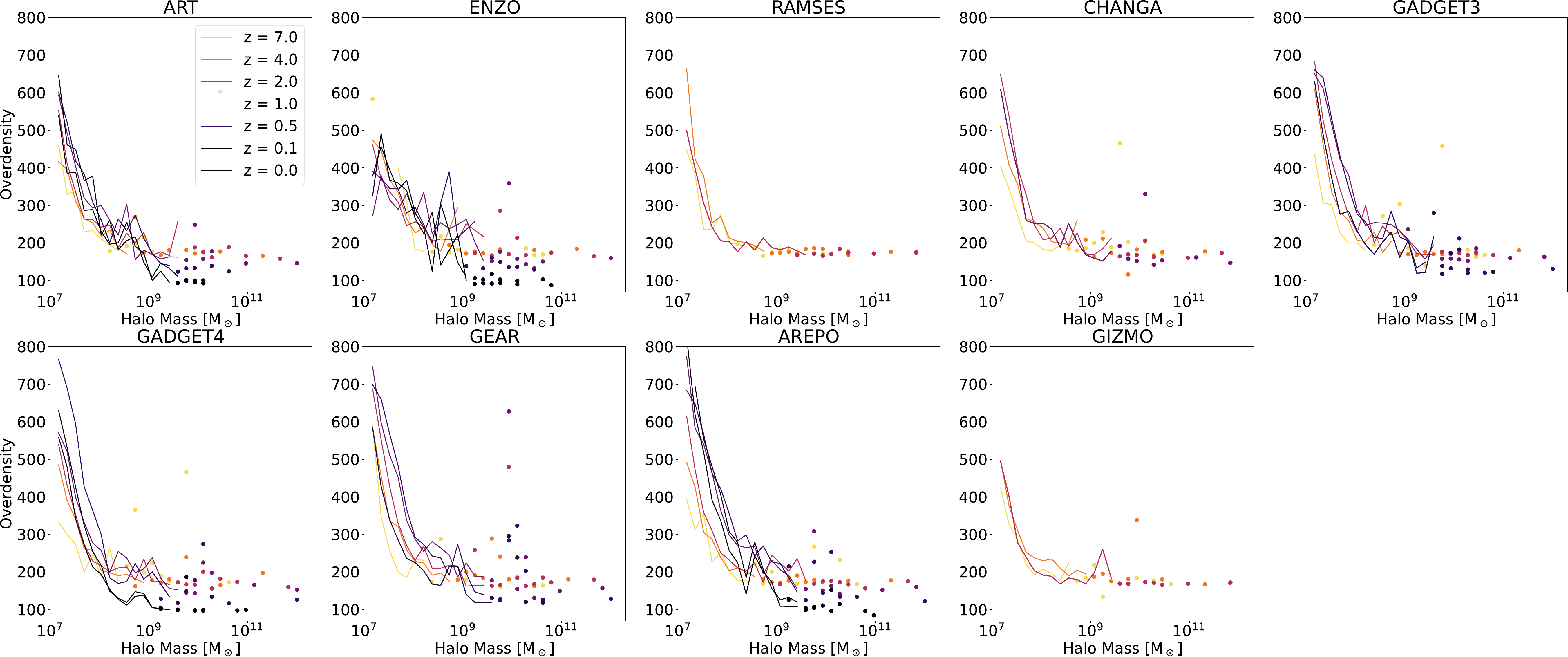}
\caption{Overdensity versus halo mass within the refined region for each of the \texttt{CosmoRun} simulations at fixed redshifts. Each data value is the mean of a mass bin. Lines are used for mass bins with greater than five halos, and scatter points are used for mass bins with five or fewer halos.}
\label{fig: allover}
\end{center}
\end{figure*}

\begin{figure}
\begin{center}
\includegraphics[width=\linewidth]{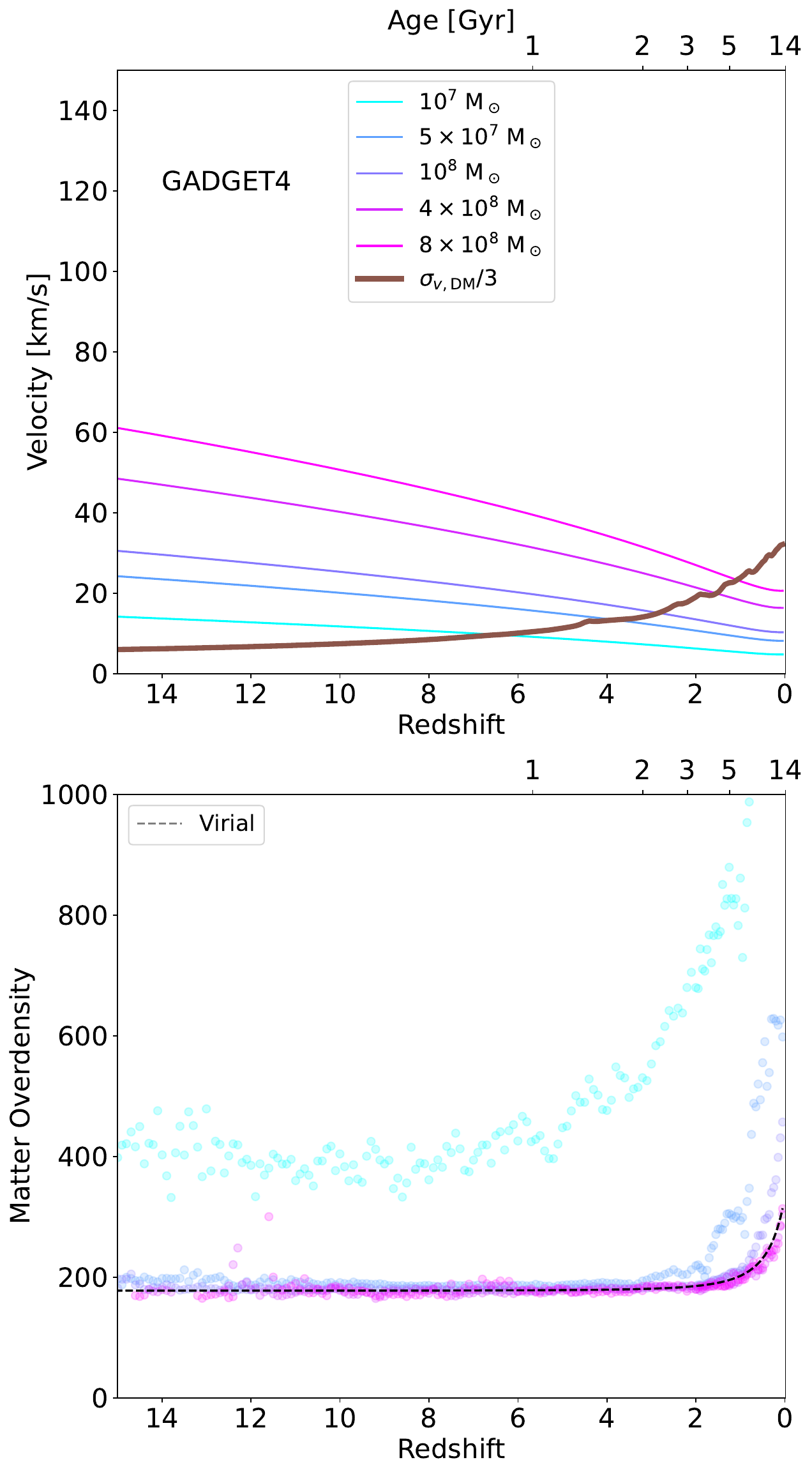}
\caption{Top: A comparison of escape velocities at the virial radius of halos of different masses (solid-colored lines) to one third the dark matter velocity dispersion (thick brown line) in the \texttt{CosmoRun} GADGET-4 simulation refined region. Bottom: Also plotted as translucent scatter points are the overdensities of halos with masses near the corresponding values used in the escape velocity calculation. The x-axis scaling emphasizes how the variables evolve with redshift rather than time.}
\label{fig: disper}
\end{center}
\end{figure}

The overdensity of the main halo is unaffected by low mass/low particle count effects even at high redshift, and so we can use its evolution as a probe of the dynamical and cosmological evolution of halos over time. Since our solving method biases halos' overdensities towards the virial overdensity over time, deviations from that value may provide insights into the redshift evolution of halo density.

Fig. \ref{fig: mainover} (top) shows the matter overdensity of the main halo (translucent line) as well as its halo mass to bound mass ratio (solid line) versus simulation time. Since the halo mass to bound mass ratio is always below unity and overdensities track the virial overdensity, the main halo is always a Class I solution in every simulation at every time step. This serves as a validation that our solving method for overdensity recovers the virial solution in the most massive, most spherical case in our data.

Like several of our other measures, the halo mass to bound mass ratio is reactive to mergers, which is expressed as a decrease in the ratio for the highest mass ratio mergers. At late times, the ratio trends upward, implying that the bound mass and the virial mass may gradually converge after $z=0$.

Fig. \ref{fig: mainover} (bottom), we explore the set of halos with mass greater than $8\times10^8$ M$_\odot$ ($\sim$1786 particles) that are between 0.5 and three virial radii of the main halo.  We set the minimum mass to avoid the effects of low particle counts on overdensity, and we set the distance to the main halo so that we can study the effect of environment and proximity to a large halo on the overdensity of halos. The values reported in these plots are simple averages of the values for the individual halos. Outside of this region, the overdensity of large halos trends towards the virial value.

As shown in the figure, before 2 Gyr, the mean overdensity scatters between the virial value and 300. Though some simulations show higher values than others at various points in time, they are roughly consistent. After 2 Gyr, the mean grows faster than the virial value, settling at around 400 for the simulations that reach $z=0$. Since the average halo mass to bound mass ratio approaches unity faster for this sample than for other halos in this mass range, the overall implication is that these halos need to be denser to maintain their integrity in the proximity of the main halo than the effect of dark energy and cosmological expansion would analytically imply. Thus, corrections to virial theorem may be needed for low redshift halos in cluster environments, including satellites of the Milky Way.

\subsection{Full Sample}

\label{sec: fullsample}

Though \texttt{CosmoRun} is based around a Milky Way-sized halo, the refined region captures other independent systems of smaller halos and creates a larger sample that we can use to probe how our measures trend more generally. However, like many studies based on zoom-in regions, we are limited by the bias of only studying an overdense region that mostly converges into the proximity of the main halo so our observations should be considered in that context.

For the full sample, the secular change in variables like sphere filling fraction and $a/b$ are the same as the trends we noted for the main halo. On average, sphere-filling is positively correlated with Halo Mass and simulation time. The higher the mass, the smaller the sample and the averages scatter. Similarly, $a/b$ is negatively correlated with halo mass and cosmic time. Though we did not find a secular $a/b$ relation with cosmic time for the main halo, jumps in $a/b$ were found to be correlated with major mergers. At higher cosmic time, the major merger rate naturally decreases, so the average $a/b$ values similarly decrease. 

The following quantities have more complex relationship, which we explore in more detail.

\subsubsection{Overdensity}

\begin{figure*}
\begin{center}
\includegraphics[width=\linewidth]{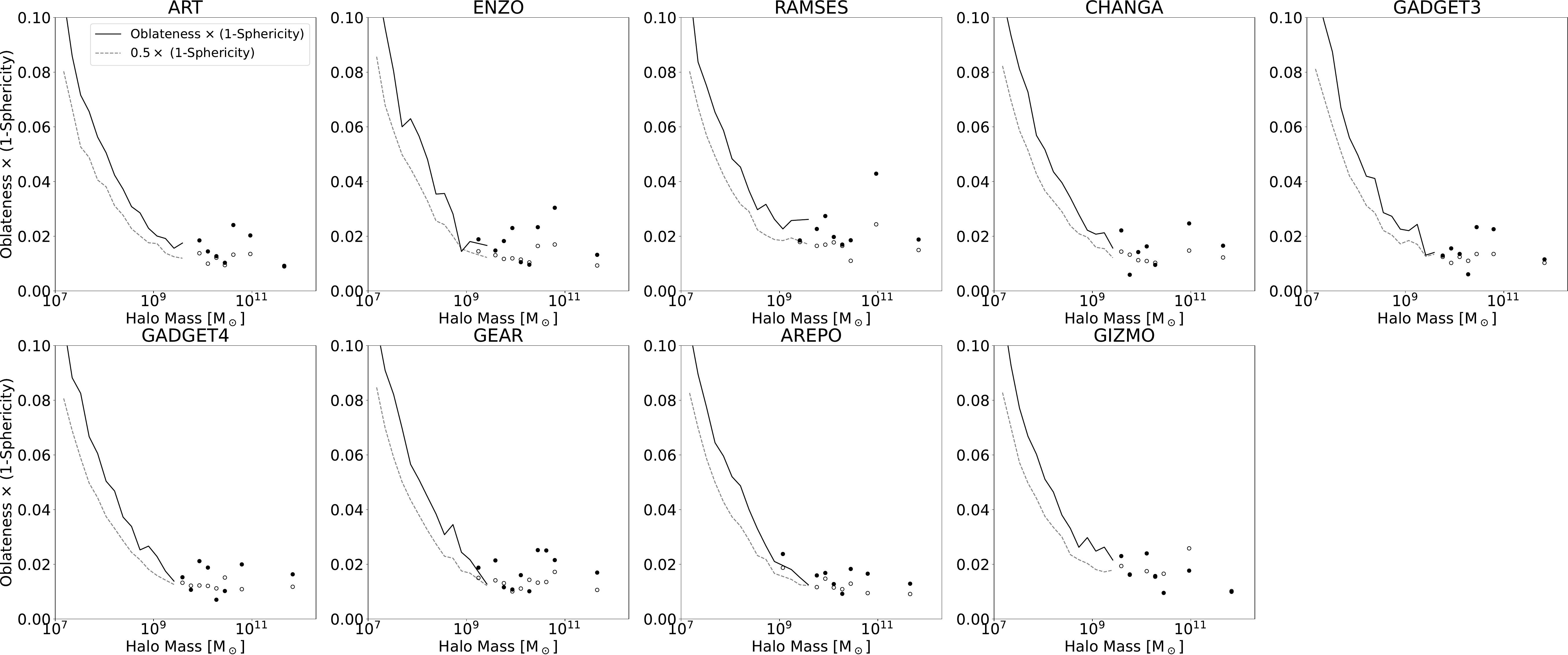}
\caption{Oblateness times (1-Sphericity), $O(1-\Psi)$, for all halos in the refined region of each simulation at $z=2$ as a function of mass (solid black line and filled circles) compared to $O(1-\Psi)/2$ (dashed line and open circles), which represents neutral oblateness. Each simulation tends towards values above the neutral oblateness line and thus is biased towards oblate orientations. In this work, this means that the semi-major axis and the angular momentum vector tend to be more perpendicular than aligned.}
\label{fig: oblate}
\end{center}
\end{figure*}

A shown in Fig. \ref{fig: allover}, the low mass relationship between overdensity and mass demonstrated in Fig. \ref{fig: combined} (top) is not independent of redshift. In all simulations, the overdensity trend from $z =7$ was lower than at lower redshifts. This implies that this relationship is being moderated by dynamics rather than particle count, as we implied in Sec. \ref{sec: overmass}. This dynamical difference can either be non-numerical (e.g., growth of particle velocities, expansion, tidal stripping) or numerical (e.g., cosmological lengthening of gravitational smoothing). As described in \citet{2021ApJ...917...64R}, after $z=9$ the particle softening length is switched from 800 co-moving pc to 80 proper pc for particle based-codes, and since the virial radius grows with time ($\sim$$3.8$ kpc for a 10$^7$ M$_\odot$ halo at $z=0$), numerical effects should dissipate over time. Therefore, the effect is likely non-numerical rather than numerical. Conversely for AMR codes (ART, ENZO, RAMSES), particle softening is fixed in co-moving units and there is a correspondingly weaker relationship between redshift and overdensity. To quantify the dynamical evolution of dark matter, we used the GADGET-4 \texttt{CosmoRun} simulation to find the dark matter velocity dispersion, $\sigma_{v,\rm{DM}}$, within the refined region over time and compared that value to the escape velocity of a halo with a virial overdensity, which is a function of halo mass. Though some of the increase in dispersion comes from dark matter particles bound to massive halos, we expect a significant fraction of the dark matter particles to fail to bind or remain bound to the halo below a threshold mass if the overall dark matter velocity dispersion is significantly higher than a halo’s escape velocity. 

In general, the escape velocity of a halo at it's outer radius with the virial overdensity ($r_{\rm vir} = (\frac{3M_{h}}{4\pi\Delta_c\rho_c})^{1/3}$) is $v_{\rm esc} = \sqrt{2GM/r_{\rm vir}}$. Thus, for a fixed mass, $v_{\rm esc}$ will tend to decrease as $p_c$ decreased with lower redshifts. Therefore $\sigma_{v,\rm{DM}}$, which increases with redshift due to cosmic expansion, will eventually be higher than a halo's escape velocity. In practice, this dark matter heating and evaporation seems to occur when the volume-wide $\sigma_{v,\rm{DM}}$ is around three times the escape velocity of a halo at the virial radius. 

As shown in Fig. \ref{fig: disper} (top), the crossing points with $\sigma_{v,\rm{DM}}/3$ (brown line) for $10^7$, $5\times 10^7$, $10^8$, $4\times 10^{8}$, and $8\times 10^8$ M$_\odot$ halos are at $z=6.5$, $z=3.8$, $z=2.8$, $z=1.5$, and $z=1.1$ respectively. Also plotted in Fig. \ref{fig: disper} (bottom) are the median matter overdensity of halos within 10\% of the indicated masses as a function of redshift (scatter points). After the escape velocity crosses $\sigma_{v,\rm{DM}}/3$, overdensities tend to rise, but only after bound particles outside the virial radius evaporate, leaving behind a denser halo.

We have confirmed that the environment can lead to higher overdensities, implying that interactions with large halos can play a role in densification, for example, through tidal stripping. Fig. \ref{fig: mainover} (bottom) showed that the mean overdensity showed a clear signal for $>8\times 10^8$ M$_\odot$ halos, the medians did not. By using the median in Fig. \ref{fig: disper}, we are effectively masking the strongest environmental effects of the main halo and other large halos.

Halos with extremely low particle counts are unable to resolve their potential wells and may need higher overdensities to bind particles to compensate. Both this limitation and a redshift dependence appear to contribute to the trends in Fig. \ref{fig: disper}. Halos with a mass of $10^7$ M$_\odot$ ($\sim$35 particles) have a higher overall matter overdensity at all redshifts to compensate for their smaller counts, which further grows after $z\sim 8$. Higher mass halos are able to track the virial overdensity (dashed black line) until lower redshifts, when their values also grow. Halo with masses $\geq4 \times 10^8$ M$_\odot$ ($\sim$1429 particles) track the virial overdensity on average at all redshifts. However, as suggested in Fig. \ref{fig: mainover}, the halo mass to bound mass ratio evolves upward with time as it becomes increasingly difficult to categorize the virial radius with a subset of the bound particles. This means that though higher mass halos are more easily characterized by a virial overdensity, the halos are still strained by cosmological effects. Conversely, since the $10^7$ M$_\odot$ halos were never characterized well by a virial overdensity, cosmological effects immediately appear that aren’t masked by our hard-coded bias towards the virial overdensity. Since this affects all mass ranges, including the highest mass halo, the underlying trend implies that the virial overdensity is not a consistent measure of a halo, even when it can be consistently calculated.

\begin{figure*}
\begin{center}
\includegraphics[width=\linewidth]{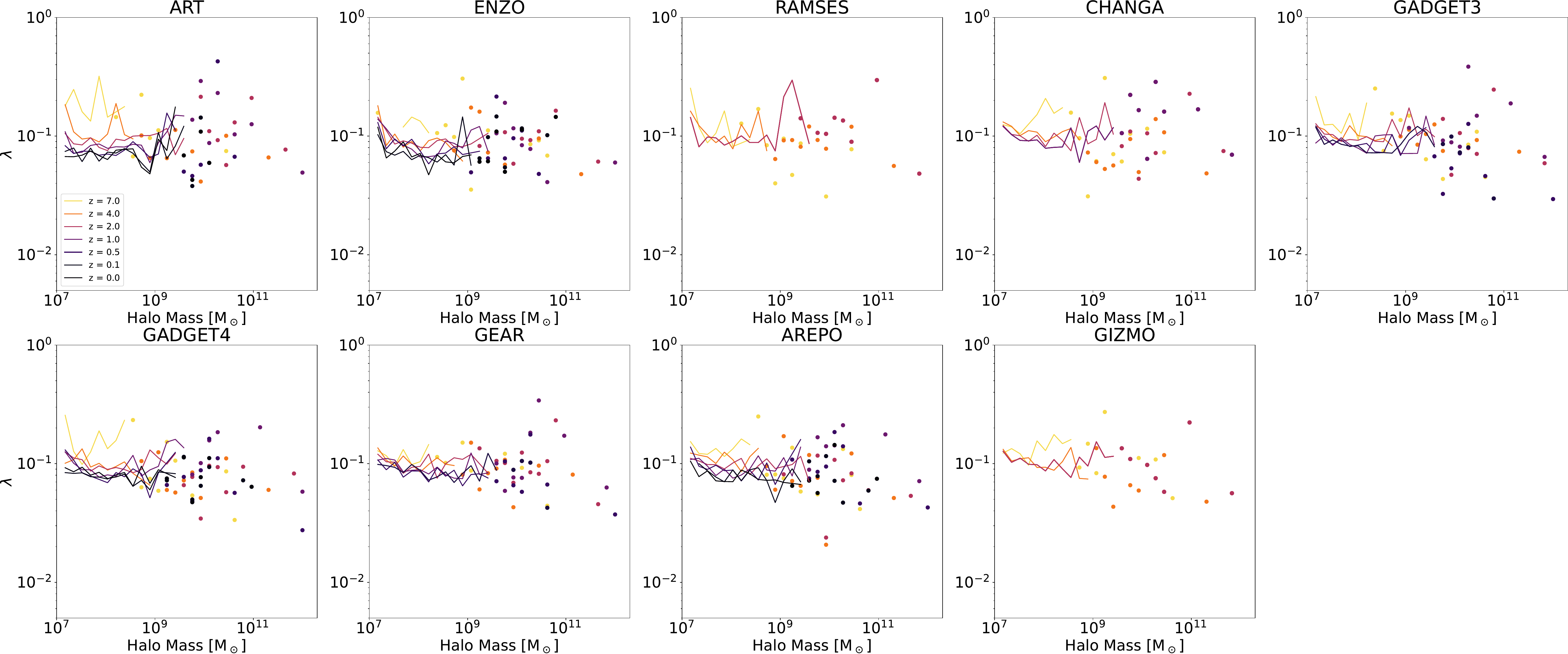}
\caption{The spin versus halo mass versus redshift trend plotted in the same manner as Fig. \ref{fig: allover}.}
\label{fig: allspin}
\end{center}
\end{figure*}

\begin{figure*}
\begin{center}
\includegraphics[width=\linewidth]{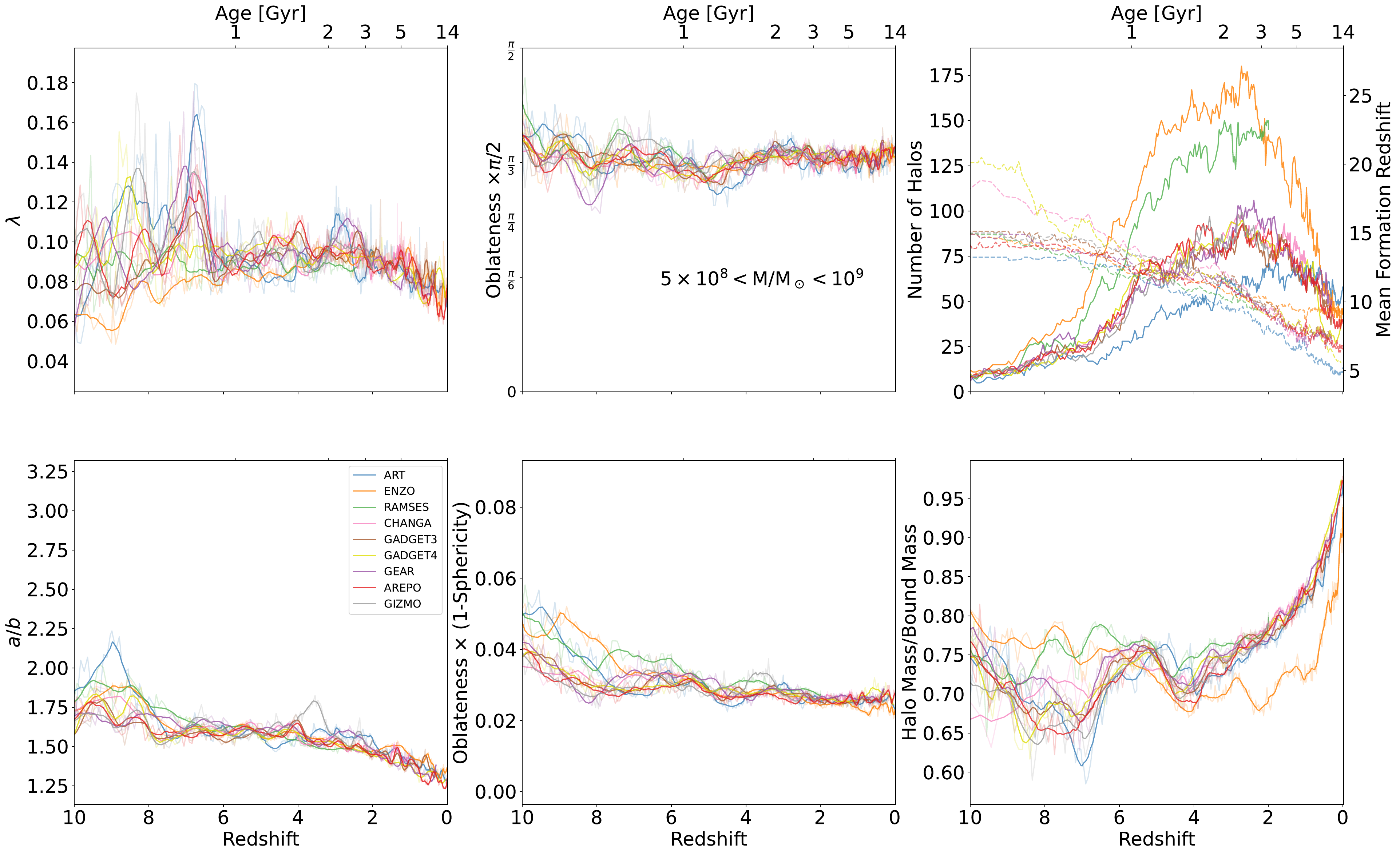}
\caption{Key dynamical and shape trends for halos in all simulations between $5 \times 10^8$ and $10^9$ M$_\odot$. In the top right plot is the sample size as a function of redshift (solid line) and the mean formation redshift of the halos in the sample as a function of redshift (dashed line). For all other plots, we show the mean values of the indicated quantities across the sample. Smoothed values are opaque and the full trend is translucent. The x-axis scaling emphasizes the how the variables evolve with redshift rather than time.}
\label{fig: spindown}
\end{center}
\end{figure*}

Extremely small halos ($10^{-7}$ M$_\odot$, Earth-sized) have been produced in cosmological simulations \citep{2005Natur.433..389D,2010ApJS..191...43S,2013JCAP...04..009A} showing their formation at high redshift as the earliest structures if enough resolution is available to resolve their density profiles. The corresponding peak densities of these halos were so high that they were explored as sources for a signal for dark matter self-annihilation. These dense cores and extremely fine particle resolutions make these ``micro-halos’’ robust against dispersion, so they can be reproduced at any redshift if one constructs enough nested zoom-ins to establish a high enough density \citep[e.g.][]{2020Natur.585...39W}. Micro-halos in these works were not identified or constrained by self-boundness, but rather by their density profiles. Using a density profile, you can construct $r_{200c}$ just by taking advantage of the smoothness of density across space and tracing density from a core to a desired overdensity. Since our halos are differently defined, they are not directly comparable.

Our smaller halos have more difficulty surviving to low redshift as the threshold overdensity growth to remain self-bounded eventually meets a numerical limit. Ordinarily, as long as enough density can be modeled to bound a particle, the central density should not affect a determination of how many particles are bound to a halo. This is because the gravitational potential is only related to enclosed mass in a spherically symmetric configuration. However, with the exception of merging halos, our smallest halos are also less spherical. As shown in Fig. \ref{fig: comp1}, the number and shape of small halos are a key difference between our spherical and non-spherical halo trees. There are several instances of small halos that only appear in the non-spherical tree, as well as non-spherical halos that are recorded as adjacent spherical halos in the spherical tree. For a highly non-spherical halo, the gravitational potential is warped, and particles are more susceptible to stripping (tidal and evaporatory) if local densities are not high enough.

\subsubsection{Oblateness and Spin}

Our definition of $O(1-\Psi)$, where $O$ (oblateness) and $\Psi$ (sphericity) are given in Eqs. \ref{eq: oblate} and \ref{eq: psi} respectively, gives us a measure of both how perpendicular the angular momentum axis is to the semi-major axis and to what degree a halo’s shape deviates from a perfect sphere. In Fig. \ref{fig: oblate}, we show the average value of $O(1-\Psi)$ as a function of halo mass at $z=2$. As we show in Fig. \ref{fig: combined} (top left), $\Psi$ is strongly positively correlated to halo mass and so $O(1-\Psi)$ is correspondingly higher for low mass halos. Since $O$ takes values between 0 and 1 linearly for angles between the semi-major axis and the angular momentum direction between 0 and $\pi/2$. $(1-\Psi)/2$ should represent neutral oblateness, wherein the angle between angular momentum vector and the semi-major axis is $\pi/2$ and the halo is neither more bulging in its poles or equator.

The average values of $O(1-\Psi)$ are systematically and consistently higher than $(1-\Psi)/2$, implying that the angular momentum axis and semi-major axis are more likely to be unaligned (oblate). This is in broad agreement with other findings \citep[e.g][]{2005ApJ...627..647B,2006ApJ...646..815S,2007MNRAS.376..215B}. However, there is a significant spread in data and individual values for $(1-\Psi)/2$ fill in the region under $(1-\Psi_{\rm min})$. Since sphericity is fairly consistent with the exception of small, short disruptions during mergers, the scatter should be attributed to large, high-frequency fluctuations in $O$ for each individual halo over time, which we observed for the main halo in each simulation (translucent lines in Fig. \ref{fig: sphereoblate}). The time averaged mean value of $O$ for the main halo was between 0.60 and 0.65 for every simulation, and so the angular momentum axis and the semi-major axis were biased toward non-alignment. In Fig. \ref{fig: sphereoblate}, $O(1-\Psi)$ (transparent lines) is seen evolving differently than  $(1-\Psi)/2$ (solid line). The halo is more oblate ($O>0.5$) during mergers as well as during the period of time shown in Fig. \ref{fig: abtrend} (bottom left) when the halo has a high offset between the halo center of mass and the hull center ($\Delta\vec{r}/r$) peaks at around 0.85 Gyr. Between these events, $O$ returns to values higher than 0.5, implying that this is the settled state. Furthermore, large changes in $O$ do not seem to directly affect $\Psi$, which is much less variable and does not move with secular changes in $O$. Dynamically, this would imply that halos torque their semi-major axis away from their angular momentum axis without changing shape.

Fig. \ref{fig: allspin} shows the relationship between mean spin, halo mass, and redshift. We see in the full sample for all simulations and at all redshifts the same general trend of spin up and down over time that we reported for the main halo. The spin values averages were generally between 0.05 and 0.2 and were scattered significantly for individual halos. As shown in Fig. \ref{fig: mainspin}, scatter was correlated to mergers, but the secular value evolved more consistently.

\section{Fixed-Mass Trends and Discussion}

\label{sec: mass-fixed}

\begin{figure}
\begin{center}
\includegraphics[width=\linewidth]{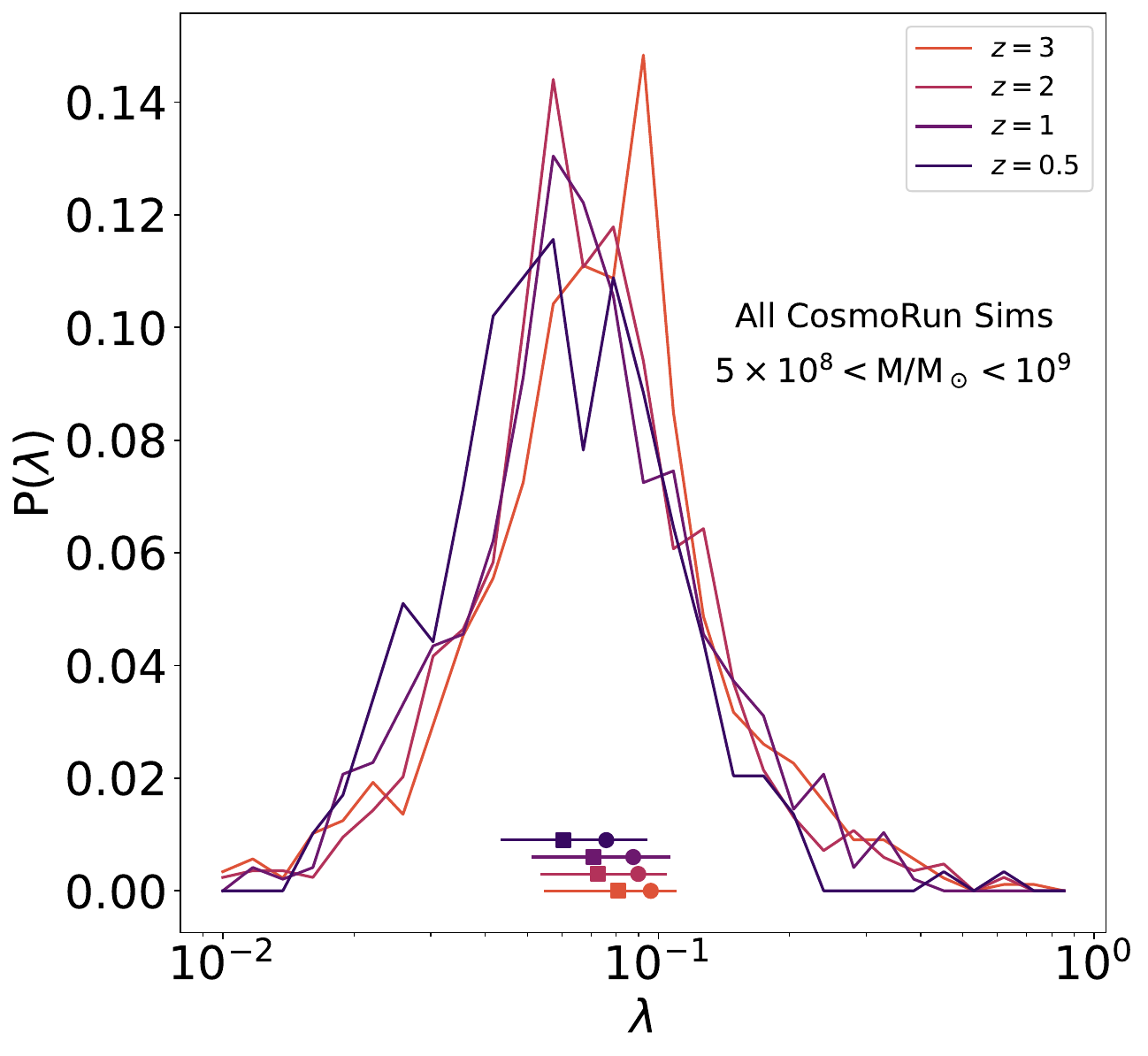}
\caption{Spin distribution constructed from the combined sample of $5 \times 10^8$ and $10^9$ M$_\odot$ halos from simulations that have data available for the given redshifts. The data show a skewed log-normal relationship with occasional bimodal peaks due to the combined sample. Circular points on the bottom show the unweighted mean of the distribution for the corresponding redshift and have a downward trend to lower spin from $z\sim3$. Box scatter points show the same trend in the median and the horizontal lines show the interquartile range.}
\label{fig: spindis}
\end{center}
\end{figure}

Several of our measures are highly responsive to the mass of the halo, so we narrowed down our sample to a smaller range of masses to study their secular evolution of with redshift. In Fig. \ref{fig: spindown}, we plotted the average values of key parameters for all halos between $5 \times 10^8$ and $10^9$ M$_\odot$ versus redshift, which is a range of masses that did not seem to be strongly affected by low particle counts or low halo count statistics for most redshifts and was consistently well-defined at the virial overdensity. The top right plot (solid lines) shows the number of halos in the sample as a function of redshift and simulation.

There were high scatter and disagreement in average spin between simulations below $z=6$ since the sample size is small, but afterwards, the means followed a tight, consistent relationship, and average spins begin to decline after $z\sim2$. Because the results are similar between simulations for $z<6$, we combined the simulation samples to build combined spin distributions for this halo mass range as shown in Fig. \ref{fig: spindis}. These distributions are roughly log-normal as indicated by previous studies, but with more pronounced skewness. Similar to the mean (shown as scatter points), the peak of the distribution trended downward with lower redshift from $z=3$. This lies in contrast to studies showing that the distribution of halo spins was largely independent of redshift \citep[e.g][]{2011MNRAS.411..584M}. Some studies have demonstrated a decrease, such as \citet{2005ApJ...634...51A}, which showed a much slighter decrease in the peak of their fitted log-normal distribution (0.003) for clusters than the (0.015-0.02) mean spin decrease we show for this mass range from $z=1$ to $z=0$. \citet{2017MNRAS.466.1625Z} showed that a similarly large decrease in the log-normal peaks of their true spin distribution could be recovered for SO halo finders. However, their trend extended to $z=8$, whereas our trend is suggestive of a peak between $z=2$ and $z=4$. Additionally, their results also showed no trend for a FoF halo finder, which is more closely aligned with our halo-finding methodology than SO finders.

There could be multiple sources for this discrepancy, such as the difference between \citet{2001ApJ...555..240B} spin and true spin, the redshift evolution of halo shape, halo mass/bound mass fraction, and overdensity, our combined sample of multiple simulation codes, or bias in the zoom-in region itself. In Fig. \ref{fig: spindown}, we present key trends that could help explain this evolution that are broadly consistent between simulations. In the top right plot (dashed line), we show that the halos in the sample are progressively younger as the older halos merge into larger halos or grow out of the mass range over time. Our single time-step results in Fig. \ref{fig: combined} (top right) show that a positive correlation between $a/b$ and our fixed mass results show a secular drop in the $a/b$ ratio and redshift in Fig. \ref{fig: spindown} (bottom left). The latter finding is the opposite of the \citet{2006MNRAS.367.1781A} finding of higher mean $a/b$ ratios with lower redshift for fixed-mass bins. This further discrepancy could be related to their $1/r^2$ weighted definition of $a/b$, which more strongly emphasizes the steepness of the density profile of halos, which is correlated with redshift through virial theorem. Our definition of an $a/b$ ratio is less sensitive to density profiles, and so the strong correlation between formation time and $a/b$ could be suggestive of a dynamical difference in late-forming halos.

Our sample also had a lower $O(1-\Psi)$ (center), but a consistent mean value of $O$, which means that sphericity ($\Psi$) increases with time. This implies that a key redshift-independent trend for a fixed-mass sample is that the average angle between the angular momentum axis and the semi-major axis, $O\times \pi/2=\theta$, is approximately $\pi/3$. Since this is neither collinear to or perpendicular to the semi-major axis, this implies a strong procession of the angular momentum vector. The distribution of cos($\theta$) in simulations have been studied \citep[e.g.][]{2006ApJ...646..815S,2007MNRAS.376..215B} showing a smooth, decreasing trend from $\cos(\theta)=1$ to $\cos(\theta)=0$. 

This sample also has an overdensity that is roughly consistent with the virial overdensity as shown in Fig. \ref{fig: disper}. However, halo mass to bound mass ratios increase with lower redshift (In Fig. \ref{fig: spindown}, bottom right), which implies a disconnect with the presumption of a relaxed virialized halo and the virial overdensity. Here we see one difference between simulations with ENZO, where there is a depression in the trend at intermediate redshift. ENZO’s refined region contains more halos in this mass range due, in part, to readjustments that were made as the main halo’s trajectory diverged out of the zoom-in regions over time. This seems to have revealed a possible environmental connection on halo mass to bound mass that did not appear in the other variables. However, because this trend does not significantly affect the other variables, those variables may not be environment-dependent or sample-dependent. The full context of all these trends will be explored with a simulation that is not biased to a zoom-in region in future work that is not focused on the \textit{AGORA} simulations.

\label{sec: allhalo}


\section{Summary}
\label{sec: summary}

We have introduced a new, non-spherical halo-finding version of {\sc Haskap Pie} that takes advantage of its open search for self-bounded particles. To test our technique, we used the \textit{AGORA} cosmological simulation comparison project's \texttt{CosmoRun} suite of simulations to study the halos we recovered with our algorithm. The dynamical and shape parameters reported by {\sc Haskap Pie} for halos include measures that do not presume that halos are symmetric or spherical, allowing an in-depth study of dark matter halo morphology.

\begin{figure*}
\begin{center}
\includegraphics[width=0.9\linewidth]{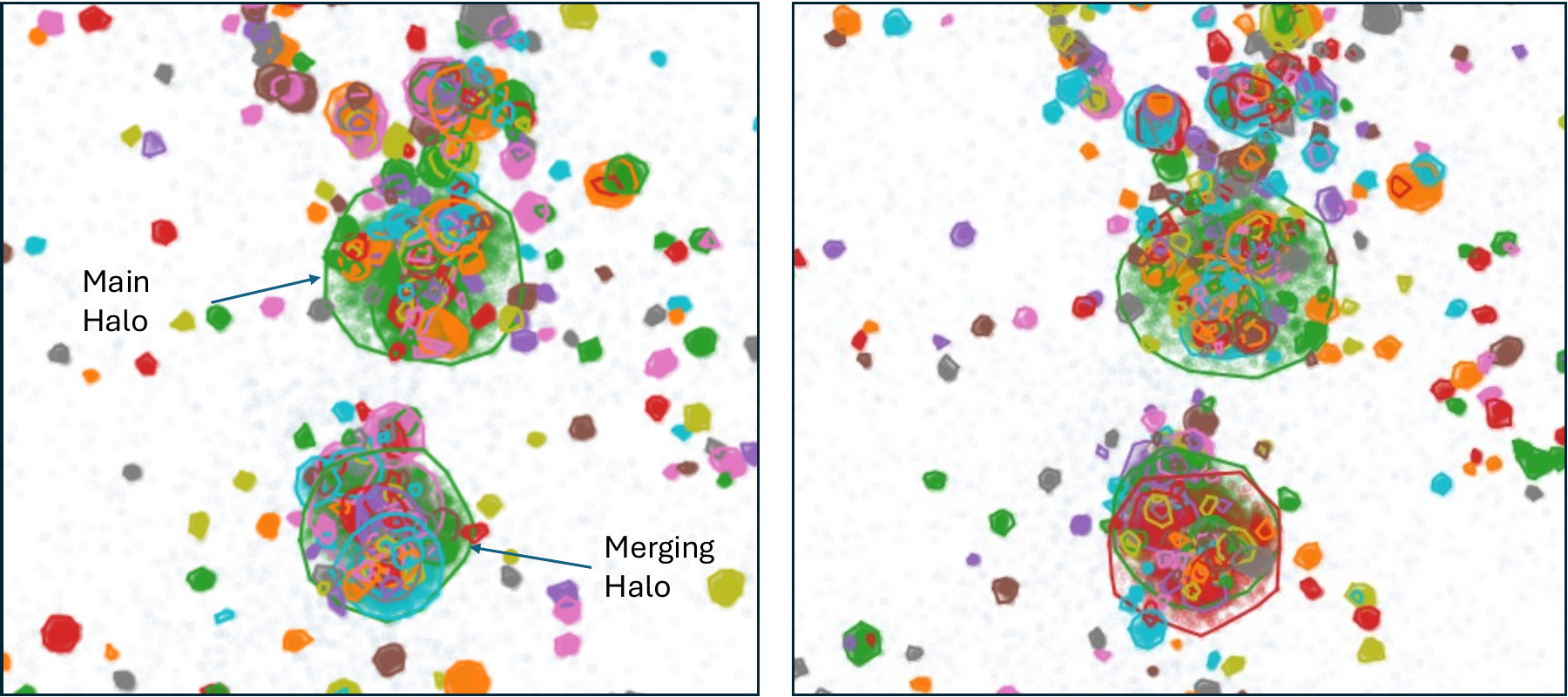}
\caption{The region around the main halo at 800 Gyr in the \texttt{CosmoRun}-CHANGA (left) and GIZMO (right) simulations showing the complex assembly of halos that make up the infalling halo for the major merger at 1.1 Gyr. This cluster of CHANGA halos loses its definition just prior to the merger, which causes the outlier mass ratio seen in Fig. \ref{fig: abtrend} (bottom) and Tab. \ref{tab:timing}. Colors are consistent by halo but rotate through the pallette between halos.}
\label{fig: changa}
\end{center}
\end{figure*}

Some of our key findings so far are:

\begin{enumerate}
    \item {\sc Haskap Pie} tracks and reports halos as an overdense collection of convex hulls that encompass self-bound particles that meet a threshold of overdensity. The resulting halos are non-spherical and not tied to density peaks. The flexibility in halo shape results in differences in the halo catalog (Fig. \ref{fig: comp1}).
    \item Broader trends for aggregate samples of halos were roughly consistent between simulation codes, but there were differences in the morphological response to major mergers tied to timing discrepancies and dynamical differences prior to mergers.
    \item Geometric parameters which include $a/b$ ratio, $\Psi$\ (Eq. \ref{eq: psi}), and $\phi$\ (Eq. \ref{eq: phi}), all respond to major mergers of the main halo. The sphere-filling fraction ($\phi$) responds more quickly to and more sharply to mergers with more intermediate mass ratios (Fig. \ref{fig: ab}). The $a/b$ ratio of our restricted sample of masses shows a secular decrease in the $a/b$ ratio from $z\sim6$ (Fig. \ref{fig: spindown} bottom left).
    \item Halo overdensity has a complex interaction with redshift, halo-mass (Fig. \ref{fig: allover}), and formation time. There appears to be a relationship between dark matter velocities and the overdensity of smaller halos (Fig. \ref{fig: disper}). The virial overdensity does not track this relationship and thus does not provide a consistent definition for our halos.
    \item Spin ($\lambda$) exhibits merger induced variations as well as secular changes for the main halo (Fig. \ref{fig: mainspin}) and the full sample (Fig. \ref{fig: allspin}). When we restrict the sample to a tighter range of masses ($5 \times 10^8$ and $10^9$ M$_\odot$) and take an average, spin peaks between $z=4$ and $z=2$ and then decreases until $z=0$ (Fig. \ref{fig: spindown} top left and Fig. \ref{fig: spindis}).
    \item The angle between the angular momentum vector and the semi-major axis is highly time varying, but for our restricted sample of masses ($5 \times 10^8$ and $10^9$ M$_\odot$), the average angle is consistently $\sim$$\pi/3$ (Fig. \ref{fig: spindown} top center). Thus, halos are generally biased towards bulging perpendicularly to their angular momentum axis but are also likely to precess.
    \item Differences in merger times between the simulations are due to actual positional and velocity differences in the location of the halo and can be predicted ahead of the merger (Fig. \ref{fig: abtrend}). Outliers in detecting the merger mass ratio of the largest merger of the main halo are due to differences in the complexity of the sub-halo configurations and gravitational potential of the infalling halo (Fig. \ref{fig: changa}). 
   
\end{enumerate}

\section{Limitations and Future Work}

\subsection{Progress with {\sc Haskap Pie}}

The non-spherical and spherical versions {\sc Haskap Pie} function as independent modes that are easily selected by the user. In the future, more dedicated algorithms for the treatment of the non-spherical case might improve the handling of some of the more unique structures and interactions we discovered in this study. For example, we have embarked on a project to better treat mergers and more consistently and correctly identify progenitors when we build the halo tree which we will shortly implement and document. Additionally, using {\sc Haskap Pie} on large datasets is prohibitively expensive and so we will consider further measures we can take to optimize the code.

\subsection{Solving Dynamical Questions}

Our analysis of the \textit{AGORA} simulations provided insights into trends that we would like to continue to explore with a larger dataset. Specifically, we will study the trends in spin and the angle between the semi-major axis and the angular momentum vector. In both cases, neither our results nor the way we measure the quantities in the context of a non-spherical halo fully agree with the literature. Since these trends were not simulation-dependent, we have identified a large single-code simulation with a large sample of halos that we will use to fully investigate this finding. That investigation is best separated from this work, which is primarily focused on introducing a new method and performing a code-comparison. Analyzing the larger sample will likely require further development of {\sc Haskap Pie}.

\section*{ACKNOWLEDGMENTS}
\begin{acknowledgments}
KSSB, THN, SM, VS, and ECS acknowledge the University of Illinois at Urbana-Champaign for their continued support. KSSB acknowledges the National Center for Supercomputing Applications as well as the Texas Advanced Supercomputing Center for their support with the Delta Supercomputer and Stampede3 Supercomputer, respectively, as well as the ACCESS program for computing grants PHY230100, PHYS240175, and PHYS250173. We acknowledge the AGORA collaboration for their support and simulation data products. We also especially acknowledge Joel Primack for his contributions in the late spring and summer of 2025 to the discussions that lead to this investigation.
\end{acknowledgments}


\appendix
\section{Multiple Major Mergers}

\label{sec: multiple}

The mass ratio of the merger into main halo in CHANGA recorded in Tab. \ref{tab:timing} was significantly smaller than it was for the other simulations.  However, in Fig. \ref{fig: mainspin} (bottom left), the secular change in mass of the main halo is the second highest. Furthermore, all the signatures of a merger in Fig. \ref{fig: ab} imply that the CHANGA merger should have been more significant. To solve this discrepancy, we investigated the merger and found that the infalling halo was solved at a higher overdensity ($\sim$388$\rho_c$) at 1000 Myr than the corresponding halo in the other simulations and that the bound mass was greater than the halo mass, making it a Class III solution at 1000 Myr whereas the others were Class I or II. This difference is connected to both the halo-finding methodology as well as dynamical differences between the simulations.

Halos are solved with both forward and backward modeling. When backward-modeling in time, merging halos are typically first captured as subhalos or infalling halos as Class II solutions within the larger potential well of the halo they merge into. As they leave the halo, their sphere of influence grows, and they become Class III solutions. Normally, they would settle into Class I solutions as the halo finder targets the virial overdensity, however if another major merger occurs, the potential well may remain complex and the minimum overdensity of the bound particles may remain constrained.

In forward-modeling, halos are typically initially discovered in isolation as Class I or II solutions, the latter of which settle into Class I solutions as particle counts grow. As forward-modeled halos merge, solutions typically convert to Class II and overdensities grow as they fall into complex potential wells where loosely bound particles are stripped. The subtle difference is that backward-modeled halos tend to have slightly higher overdensities before a merger than forward-modeled ones. When pruning the final solution between realizations of a halo track, the halo-finder prioritizes the solution with the longer halo history and the higher cumulative mass, which typically favors the forward-modeled solution as long as the halo paths are unbroken in each direction.

However, there are dynamical situations that have a higher tendency of breaking the lower-overdensity forward-modeled paths. Particularly, when major mergers occur in rapid succession and the bound mass rapidly changes. For example, during the infall of multiple halos, particles may temporarily unbind from the center of energy. This can lead to disfavored discontinuities in the combination of halo mass, radius, and overdensity between timesteps that fail the cost functions for the lower overdensity solution but remain consistent for the higher overdensity solution that is less sensitive to changes in the overall potential well.

CHANGA’s halo is unique among the simulations as measured by the number of major halos in its vicinity. As shown in Fig. \ref{fig: changa}, there are at least 11  massive halos (at least 5\% of the mass of the main halo) that are actively interacting with and within 10 kpc of the merging halo at 800 Myr, which ranged from two to nine for the other simulations. The encompassing halo of the complex (large green halo to the bottom in Fig. \ref{fig: changa}) buckles by 1000 Myr, leaving behind a collection of smaller halos that individually fall into the main halo. Taking the total mass ratio of halos to the main halo that are within 5 kpc of the merging halo at 1000 Myr (recorded as the ``Cluster Ratio’’ in Tab. \ref{tab:timing}), CHANGA is no longer an outlier in its relationship between merging mass ratio and peak $a/b$. However, that measure is not robust against double-counting particles, so we do not use it in our primary analysis. The merging halos in AREPO and GADGET4 also have high numbers of massive interacting halos within 10 kpc at 800 Myr (seven and nine, respectively) but have not lost the encompassing halo by 1000 Myr. Since their cluster ratios include both the encompassing halo and its components, it is significantly higher.

We have named the phenomenon of an encompassing halo that bounds a region of several distinct self-bound halos of similar mass without a clear central halo a ``cluster halo’'. On a smaller scale, the halo depicted in Fig. \ref{fig: overrad} is also a cluster halo as three separately tracked major halos have combined around a halo centered between their individual centers to form a common, distinct potential well. N-body interactions between similar masses in close proximity are chaotic and do not conserve orbital energy with respect to the center of mass. Because we have defined halos as being unique self-bound structures rather than an overdense collection of dark matter, these clusters can become unstable in rapid changes to the potential well, such as the in the CHANGA merging halo shown in Fig. \ref{fig: changa}. In this version of {\sc Haskap Pie}, this behavior mostly arises from the switch from spherical to non-spherical halo definitions and it seems to at least partially arise from unexplored dynamical considerations. Therefore, though it complicates the analysis of halos, we have not changed our finding algorithm to specifically disfavor this phenomenon so that it can be properly studied in this and future work.





\bibliography{main}
\bibliographystyle{aasjournal}

\end{document}